# Diffusion of tacit knowledge in a company: a mathematical model based on diffusion on graphs

Radosław A. Kycia[1], Agnieszka Niemczynowicz[2], Andrzej Buszko[3]

[1]Faculty of Computer Science and Telecommunications, Cracow University of Technology, Poland;

kycia.radoslaw@gmail.com; ORCID: https://orcid.org/0000-0002-6390-4627

[2]Faculty of Computer Science and Telecommunications, Cracow University of Technology, Poland;

a.niemczynowicz@pk.edu.pl; ORCID: https://orcid.org/0000-0002-4370-3326

[3]Pomeranian University in Słupsk, Słupsk, Poland; buszko@uwm.edu.pl; ORCID:

https://orcid.org/0000-0003-0600-4646

## Abstract

This article is devoted to the process of diffusing tacit knowledge. This intangible asset has proven crucial for achieving a competitive advantage among market-oriented companies. A novel model of tacit knowledge diffusion is presented, employing the concept of heat diffusion from physics. Furthermore, graph theory and the dynamics it defines are utilized. We defined a Tacit Knowledge Transfer Graph that encodes data from questionnaires. It enables us to identify employees' learning needs, allowing for the planning of classes within the same period, such as a day, to schedule them optimally. Moreover, the model can be used to identify employees with high knowledge demands. The application is not limited to companies; a simple example of planning classes in a school/university is provided. The presented model can help with optimal scheduling, enhance operational efficiency, and improve talent management within the company. It can also identify risks associated with critical sources of knowledge and help improve organizational culture and knowledge management policy.

## 1. Introduction

Even though intangible assets play a crucial role in a company's performance, particularly in a market-oriented economy, tacit knowledge transfer has been widely acknowledged as a critical driver of organizational learning, innovation, and sustained competitive advantage. Much of the prior research remains conceptually oriented, focusing on defining tacit knowledge and explaining why it is difficult to articulate or codify. While these studies provide valuable theoretical grounding, empirical evidence on how tacit knowledge is actually transferred in real organizational settings remains limited. In particular, there is a lack of understanding of the micro-level processes -such as informal interactions, shared practices, mentoring, and socialization - through which tacit knowledge is transferred between individuals and groups. The great challenge of today's companies is not only to acquire information and transform it into valuable knowledge assets, but also to explore the mechanisms of transferring tacit knowledge. Such a problem occurs in both big and small or medium companies (Rosselini, Hawamdeh, 2020). There are methodological limitations that persist in the literature. Many studies rely on cross-sectional designs and self-reported measures, which may fail to capture the dynamic, longitudinal, and often unconscious nature of tacit knowledge transfer. This creates a need for more qualitative, longitudinal, and mixed-method approaches that can better uncover how tacit knowledge is shared, retained, or lost over time. Seminal works have established the foundational distinction between tacit and explicit knowledge, emphasizing the inherently personal, experience-based, and context-specific nature of tacit knowledge. Despite this theoretical recognition, the existing literature reveals several persistent gaps that justify further investigation. Although knowledge management is generally associated with tacit knowledge, the availability of suitable tools to uncover it remains a significant challenge. Some tools can enhance this process, while others may prove ineffective or even useless.

Tacit knowledge can be viewed from two crucial perspectives: theoretical and practical. The theoretical perspective makes assumptions about and explores the impact of hypothetical developments, while the practical perspective verifies established procedures and demonstrates the clear advantages of applying new tacit knowledge. These two perspectives are closely connected and should be considered complementary assets, existing not only within the company but also within its business market niche.

A significant challenge is to investigate not only the level of tacit knowledge within a company but also how employees possessing these assets collaborate. This collaboration generates synergistic effects. Employees with a high level of tacit knowledge should work closely together to activate this knowledge. From a management perspective, it is essential to utilize all company resources effectively. Underutilization of assets leads to company losses, particularly in terms of effectiveness.

Keepers of tacit knowledge, though often evaluated on their performance, are frequently unaware of its inherent value. For example, an excellent athlete, such as a runner, may perform well but struggle to explain their performance scientifically (e.g., biomechanics). Considering the two important factors of individual performance and employee cooperation, it is essential to determine how mutual support can enhance overall organizational performance. Activating and sharing tacit knowledge is crucial for its disclosure and application. Daga, Srivastava, and Choudhary (2025) point out that the lack of tacit knowledge transfer is a key factor in reducing company performance.

Managers strive to adopt effective tools for knowledge management, particularly for leveraging tacit resources, to gain a competitive advantage. The construction industry plays a crucial role in any economy, accounting for approximately 10-12% of GDP and generating a substantial number of jobs while stimulating other sectors. In a construction company, tacit knowledge is vital because many project phases rely on experience and trial-and-error learning. This knowledge is personal, context-specific, and distributed among all workers, making its disclosure and application in construction management challenging. This article aims to fill a research gap by presenting a method for disclosing and transferring tacit knowledge, employing a cognitive map. This objective challenges the view of some researchers who claim that tacit knowledge is a purely personal asset that is hard to share (Polanyi, 1966). On the other hand, others are more flexible in sharing tacit knowledge (Gill, 2000).

Historically, professional knowledge was highly valued and guarded. In the Middle Ages, for example, according to London's Plea Roll of 1355, all craftsmen and 'people of secret' (those possessing professional knowledge) were obliged to group into guilds that ensured the secrets of their professions were kept safe (see, e.g., Richardson (2004) for a detailed exposition). However, the Renaissance and the later development of professional and technical education systems increasingly transformed guilds into interest groups capable of lobbying for the business interests of specific professions. From a modern company's perspective, however, the adequate circulation of knowledge in the workplace enhances worker performance and can accelerate work progress. Existing studies tend to privilege formal mechanisms of knowledge transfer, such as documentation systems, training programs, and information technologies. This emphasis is misaligned with the nature of tacit knowledge, which often resists formalization.

The optimal planning of classes is referred to as the UCTP (University Course Timetable Problem), and it is an NP-complete (non-deterministic polynomial) problem. It belongs to the class of problems that are difficult to solve, and typically, some heuristic algorithms are used. Therefore, every additional algorithm and insight that can speed up class planning is important. A detailed review of this discipline is presented in (Mühlenthaler, 2015).

Accordingly, this study should be understood as a theoretical and model-building contribution that focuses on the formalization of tacit knowledge diffusion on organizational networks. This research occupies a distinct niche at the intersection of knowledge management, organizational theory, and network science by advancing a theoretical and model-building contribution that formally represents the diffusion of tacit knowledge within organizational networks.

Existing literature on knowledge diffusion has predominantly focused on explicit, codified knowledge—often modelled through information transmission, learning curves, or technology adoption frameworks—tacit knowledge remains under-theorized and under-formalized. Tacit knowledge is inherently experiential, relational, and context-dependent, making it resistant to traditional analytical and quantitative modelling approaches. As a result, much of the current understanding is descriptive, metaphorical, or empirically fragmented.

This research addresses this gap by developing formal models that capture the unique properties of tacit knowledge diffusion, especially in construction business. The niche is distinguished by shifting the analytical focus from knowledge as a transferable object to knowledge as an emergent process, embedded in social relations and organizational routines. By explicitly modelling how tacit knowledge diffuses through networks—rather than assuming it spreads analogously to information—this research contributes a conceptual and mathematical formalization that is currently missing from both knowledge management and organizational network theory.

Empirical data collection and validation are deliberately positioned outside the scope of the present paper.

This manuscript is organized as follows. The first part is devoted to the introduction and literature review. This is followed by methodological assumptions, the model of tacit knowledge and its flow, and a discussion. Finally, conclusions are presented.

## 2. Literature review

Common beliefs, organizational culture, and methods for solving problems and avoiding risk can be understood as factors that provide collaborative advantages (Bergman, et all. 2016). Knowledge management is an important strategy in the construction industry. It is primarily linked to individual skills, which in turn shape project execution. However, transferring individual knowledge to teams is difficult. It can fail for numerous reasons, including a lack of proper tools for both knowledge disclosure and its conveyance to the appropriate workers (Kwong, Lee, 2009). To succeed in the construction business, project targets must be linked not only to physical execution but also to financial outcomes.

Jafari, Akhavan, Nikookar, (2013) revealed that personal knowledge is associated with corporate competencies. When personal knowledge is well-recognized and integrated within an organization, corporate competencies become more useful in daily performance. Unique competencies, in fact, rely on tacit knowledge (Moustaghfir, Schiuma, 2013). Haldin-Herrgard (2000) studied the disclosure of tacit knowledge. Controlled knowledge is often disseminated through diverse techniques, not only for detection but also for storage and sharing.

Even though tacit knowledge is recognized as a category, there is limited information on its quantity and the way it spreads within the organization (Brockmann, Anthony, 1998). It is challenging to utilize tacit knowledge within the framework of the entire organization, since it is essentially related to the personal entity (Augier, Vendelo, 1999). So, if the tacit knowledge can be regarded as relatively high at the company, but spread among individual persons – not exploited by the entire organization – it is a wrong attitude for optimal asset employment. From the effectiveness standpoint, it means a harmful strategy. It should have been remembered that tacit knowledge is the key driver of innovation (Leonard, Sensiper, 1998). Although tacit knowledge is sometimes ignored or neglected by managers, there are experiences of transferring it to the organization. Powell (2023) demonstrated that, under specific circumstances, tacit knowledge can be exploited. Such a transfer was observed during a nepotistic hiring practice. It was the case of employing river pilots. Nepotistic practices are generally blamed, but they can facilitate tacit knowledge sharing. So, whenever the benefits of exploring tacit knowledge are high, the organization can incur losses from nepotism. This way of exploration was developed by Holste and Fields (2010), as well. They raise the problem of trust in tacit knowledge sharing. When there is high confidence, both affect-based and cognition-based trust lead workers to be more willing to share tacit knowledge.

Boamah, et. all (2023) focused on the construction industry and found that tacit knowledge sharing boosts firm productivity. In this sector, tacit knowledge is gained mainly through experience and practice. Explicit knowledge must be verified on the construction site, where modifications are often required, thereby providing an impetus for the development of tacit knowledge. Chen and Mohamed (2010) investigated construction companies in Hong Kong and found that tacit knowledge is acquired individually and is a valuable asset to the company. Whenever managers introduced a knowledge management strategy, they paid close attention to worker cooperation, and tacit knowledge was transferred accordingly. They could not provide a detailed account of this transformation, but the effects on company performance were very positive.

Endres, et. all. (2007) highlighted the way of possible tacit knowledge sharing. They proposed the self-efficacy model in flexible organizations. Whenever the workers are appropriately motivated, the willingness to employ and transfer tacit knowledge is higher. The stimulus process is supposed to be based not only on self–efficacy attitudes but also on an open-society environment. Managers should take into account employees' preferences regarding team creation. In friendly surroundings workers are expected to cooperate more eagerly and share their experience. Furthermore, they can provide original solutions to solve the problem. Construction project execution can be viewed from several perspectives, and creative proposals focused on effectiveness, quality, logistics, and even safety are generally welcomed at the construction site. Suggestions can be based on experience and tacit knowledge. Gamble (2020) offered an interesting view on both tacit and explicit knowledge. Those management categories exist together in organization structure. The rise of explicit knowledge can be supported by the disclosure of tacit knowledge. This process may be observed within or outside an organization. Pathirage, et.all. (2007) confirmed a relatively high level of tacit knowledge in construction companies. Workers' involvement in project realization influences the final effect. The awareness that people are the key asset; managers are supposed to create conditions to employ all abilities of construction workers, including their tacit knowledge. Shi et al. (2022) studied collaborative innovation activities in the construction industry based on the employment of explicit and tacit knowledge. They found that explicit knowledge sharing does not have a positive effect on innovation, but tacit knowledge has a progressive impact on organizational performance. Such conclusions are significant, as they highlight the important role of tacit knowledge in construction companies.

Another notion related to knowledge is the diffusion of knowledge/innovation. Research on the use of physical diffusion models for innovation diffusion was initiated by Rogers (2003). The difference between tacit knowledge and the diffusion of innovations lies at their core: the former concerns personal knowing, while the latter concerns a social process.

The history of the application of concepts from physical theories to Economics has a long history. That includes the gravity model of trade (Isard, 1954), the whole direction of Econophysics, e.g., (Rosario, Eugene, 2007), or applications of Thermodynamics to Economy in (Kümmel, 2011). Even more advanced fundamental theories of physics, like gauge theories, are useful in the Economy, see, e.g., (Ilinski, 2001). The Fourier law describes the transfer of heat in a rod made of some material between two endpoints at different temperatures; i.e., a temperature gradient induces heat transfer, see (Cengel and Ghajar, 2019). This idea can be extrapolated to the knowledge gradients defined below.

**3. Research assumptions and discussion**

This research aims to create a model of tacit knowledge and its transfer. Multidimensional matrices are proposed as the foundation of the model to reflect the team collaboration of workers and the sharing of tacit knowledge at the workplace, such as on a construction site. Project implementation relies on teams of workers from different companies or specializations (see Figure 1).

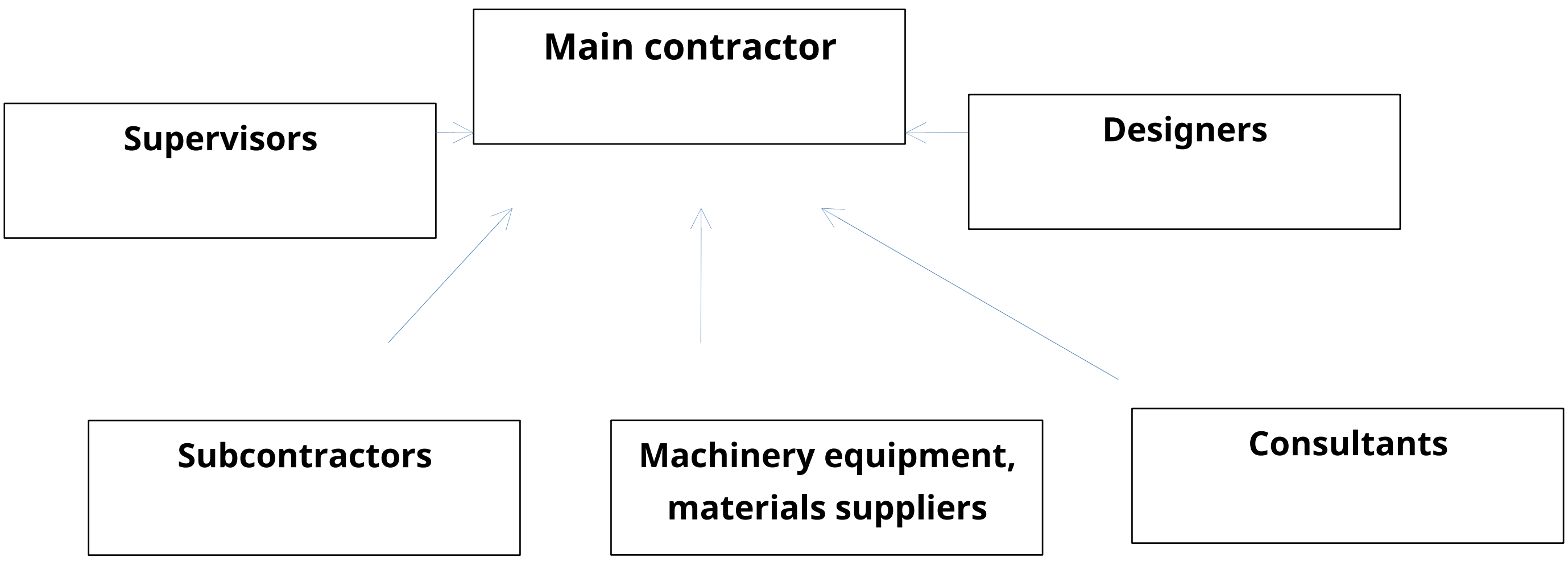

**Figure 1** The relations among companies involved in construction project.

Although there is one main contractor, there are also subcontractors, supervision inspectors, safety services, and designers. This requires teams of workers from all involved companies who collaborate constantly.

**4. The model**

The preparation of model begins by associating the set of specialists $P=\{p_1, p_2, \ldots\}$ to specific domains of knowledge: $D=\{d_1, d_2, \ldots\}$. We can also define a function $d: P \rightarrow D$, that associates an employee $p_i$ with their domain $d(p_i)= d_k$. The model is based on surveys designed to answer questions about the need for, or excess of, tacit knowledge among employees. The detailed questions depend on the specifics of the company but should be sufficient to quantify the parameters described below. The guidelines for preparing the questionnaire is presented in the subsection Data collection and operationalization below.

Each employee's interaction with other employee in the model is represented by a tuple of numbers which describes:

**a - Excess or shortage of tacit knowledge**: a value in the range $[-1,1]$. This is a function of employees $p_i$, $p_j$, i.e., $a = a_{ij} = a(p_i, p_j)$. The first index i represents the 'destination of the knowledge flow', and the second index, 'j', the source of knowledge. An excess is represented by a positive number, and a demand/shortage by a negative number. Any measurement scale from questionnaires can be normalized to this range. We distinguish two cases:

- $a_{ij}(1) \geq 0$, when the $p_j$ wants to transfer knowledge to $p_i$ – knowledge excess (source);
- $a_{ij}(2) \leq 0$, when $p_i$ needs knowledge form $p_j$ – knowledge deficiency (sink);

**b - Propensity to share knowledge**: A value in the range [0,1]. This represents an individual's tendency to share knowledge with another employee and is a function of the giver $p_i$ and the taker $p_j$, i.e., $b_{ij}=b(p_i,p_j)$. This tendency depends strongly on psychological and sociological factors. The convention of indices is the same as for $a_{ij}$. As before, we must distinguish two cases:

- $b_{ij}(1)$ – represents the propensity to share knowledge from $p_j$ to $p_i$ as measured by $p_j$; We assume that if $p_j$ likes to work with $p_i$ during teaching then also $p_j$ will like to work with $p_i$ during learning, i.e., $b_{ij}(1)=b_{ji}(2)$. This may be not true if $p_i$ is e.g., good student but not a good teacher.

- $b_{ij}(2)$ - represents the propensity of knowledge share from $p_j$ to $p_i$ as measured by $p_i$; Likewise we assume, for simplicity that $b_{ij}(2)=b_{ji}(1)$, which also might be not true when we have differences in teaching and studying abilities.

**$d_i$ - The specialist's domain**: The domain from which the specialist originates. It is an integer number describing the domain.

These tuples are organized in an array (Table 1), where rows and columns are indexed by specialists. The diagonal elements are zero, as they would describe knowledge transfer within the same employee.

| TO \ FROM | **$p_1$** | **$p_2$** | **$p_3$** | **$p_4$** |
|---|---|---|---|---|
| **$p_1$** | 0 | [$a_{12}(1)$, $a_{12}(2)$, $b_{12}(1)$, $b_{12}(2)$] | [$a_{13}(1)$, $a_{13}(2)$, $b_{13}(1)$, $b_{13}(2)$] | [$a_{14}(1)$, $a_{14}(2)$, $b_{14}(1)$, $b_{14}(2)$] |
| **$p_2$** | [$a_{21}(1)$, $a_{21}(2)$, $b_{21}(1)$, $b_{21}(2)$] | 0 | [$a_{23}(1)$, $a_{23}(2)$, $b_{23}(1)$, $b_{23}(2)$] | [$a_{24}(1)$, $a_{24}(2)$, $b_{24}(1)$, $b_{24}(2)$] |
| **$p_3$** | [$a_{31}(1)$, $a_{31}(2)$, $b_{31}(1)$, $b_{31}(2)$] | [$a_{32}(1)$, $a_{32}(2)$, $b_{32}(1)$, $b_{32}(2)$] | 0 | [$a_{34}(1)$, $a_{34}(2)$, $b_{34}(1)$, $b_{34}(2)$] |

| $p_4$ | $[a_{41}(1), a_{41}(2), b_{41}(1), b_{41}(2)]$ | $[a_{42}(1), a_{42}(2), b_{42}(1), b_{42}(2)]$ | $[a_{43}(1), a_{43}(2), b_{43}(1), b_{43}(2)]$ | 0 |
|---|---|---|---|---|

**Table 1.** The way of organizing the data for the model. Shaded part represents the values (tuples) and non-shaded part is only for indexing of this two-dimensional matrix. The domains are implicitly given by the employee $p_i$.

When there is also need to transfer the knowledge from the actor (an external agent) then we must also assign it as an employee to have the whole picture of knowledge transfers.

If an employee represents the knowledge from various domains, then to make the model consistent the employee should be assigned under a new index assigned to the next domain of expertise. Therefore the proposed model can handle specialists with multidisciplinary knowledge.

Since the situation within a company changes over time, the coefficients in Table 1., (a,b) are time-dependent and can be updated by repeating the surveys over time. For this analysis, we will focus on a specific instance in time forming a static model.

The Table 1, which we call **W**, can be seen as an array of vectors, or a multidimensional array[1]. To construct this matrix, one must:

1. Identify domains of knowledge in the company: $D= \{d_1, d_2, d_3,...\}$

2. Identify all relevant employees and actors: $P = \{p_1, p_2, p_3,\dots\}$

[1] Such tables are sometimes called tensors, however, in the precise meanin, see e.g., Penrose, Rindler, (1987) or Ruíz-Tolosa, Castillo, (2005), the tensors are mutlidimensional matrices with specific transformation laws attached. In this case there is no obvious transformation laws associated to all the types of variable a and b.

3. Using questionnaires, survey the demand or excess of the knowledge ($a_{ij}$) and the willingness to share the knowledge $b_{ij}$ between employees $p_i$ and $p_j$.

4. Construct the table **A**.

### 4.1. Data collection and operationalization

This subsection outlines the method for preparing the questionnaire. Feedback can be collected using a Likert scale (e.g., 1 – Strongly Disagree, to 5 – Strongly Agree) and then rescaled to the desired ranges.

For parameters **a** (excess or shortage of tacit knowledge) the example questions for $p_i$ about $p_j$ and domain $d_j$:

**Knowledge offer:** "I feel confident that I could effectively mentor $p_j$ on the basics of my expertise needed for their work." A strong positive response yields a value $a_{ij}(1)$ and indicates a high positive value of $a_{ij}$.

**Knowledge requests:** "When facing a complex challenge in my domain ($d_i$), I believe I would need significant guidance from $p_j$." A strong positive response yields a value $a_{ij}(2)$ and indicates a high negative value of $a_{ij}$.

For parameters **b** (propensity to share), which are linked to interpersonal trust, example questions for $p_i$ about $p_j$ could be:

- For $B_{ij}(1)$ – The employee $p_j$: "I am willing to dedicate my time to help $p_i$ solve a work-related problem, even if it is not directly my responsibility." or "I feel that $p_j$ is receptive and appreciative when I offer professional advice."

- For $B_{ij}(2)$ – The employee $p_i$: "I am willing to learn from $p_j$ to solve a work-related problem."

The answer $B_{ij}$ from the Likert scale can be rescaled to $b_{ij}$ using the simple formula:

$$b_{ij} = \frac{B_{ij}}{5},$$

where 5 indicates the maximal value of the used Likert scale.

The values $a_{ij}$ and $b_{ij}$ can be influenced by social desirability bias, see Krumpal (2013), and the Dunning-Kruger effect, see Kruger and Dunning (1999). To mitigate these biases, one can use anonymous questionnaires and average the answers from an employee and their manager, or, more expensively, use 360-degree feedback involving the employee, their subordinates, and supervisors.

### 4.2. Tacit Knowledge Transfer Graph

The matrix **A** can be used to optimize tacit knowledge flow in the company. To this end we use the analogy with Fourier Law of heat conduction[2]. We use this idea only to motivate the introduction of a gradient that describes a possible imbalance of knowledge "potentials" from a specific discipline between two employees and the direction of its flow when suitable conditions in the company are organized.

In order to motivate the connection, fix i and j, $i \neq j$, and observe that the product $V_1 = a_{ij}(1)b_{ij}(1)$ is a positive number that describes the 'potential' for knowledge transfer from expert $p_j$ to $p_i$ of domains $d(p_j)$ and $d(p_i)$ respectively. Likewise, the product $V_2 = a_{ij}(2)b_{ij}(2)$ is a negative number and represents the potential of 'deficiency' in knowledge of $p_i$ with respect to $p_j$ seen as a source of knowledge. Intuitively, the higher the values of $|V_1|$ and $|V_2|$, the higher match between *the knowledge source* (an employee who offers the knowledge) $p_j$ and *the knowledge sink* (an employee who needs the knowledge) $p_j$. This observation motivates the simplest construction of the gradient of knowledge from $p_j$ to $p_i$ as

$$\nabla a_{ij} = \frac{a_{ij}(1) b_{ij}(1) - a_{ji}(2) b_{ji}(2)}{2}.$$

[2] Fourier's Law states that the rate of heat flow (or heat flux) through a material is proportional to the temperature gradient across it and to its thermal conductivity.

One can easily observe that the gradient is non-negative and scaled to $\nabla a_{ij} \in [0,1]$. One can construct the gradient in the opposite direction $\nabla a_{ji}$. Both of these gradients are connecting vertices $p_i$ and $p_j$ however in different directions, as presented in Figure 2.

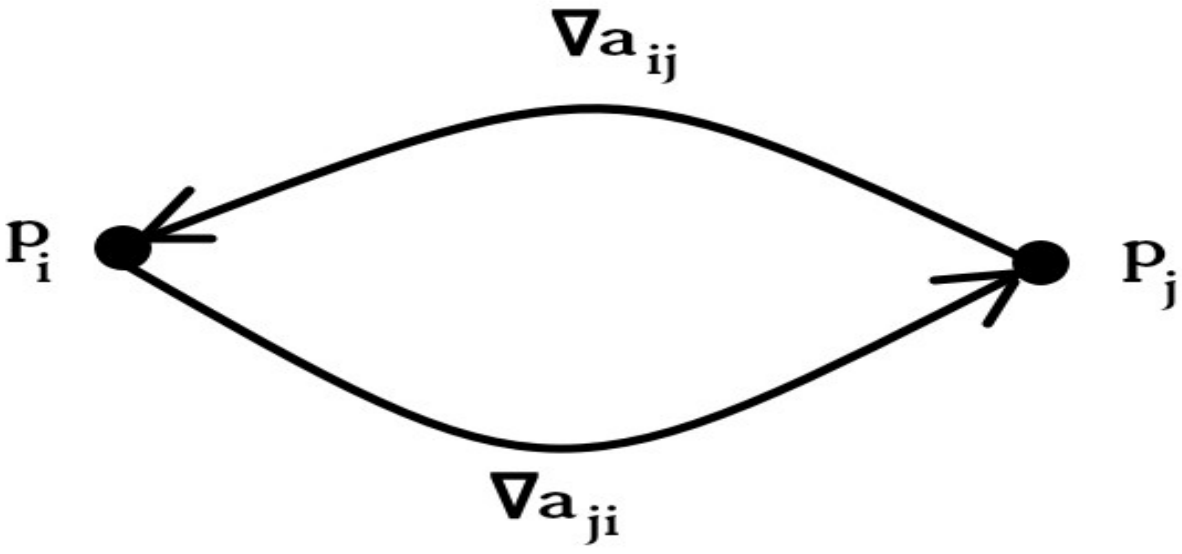

**Figure 2.** The knowledge transfer relation between employees i and j.

The construction is a part of a directed graph called a T**acit Knowledge Transfer Graph (TKTG)** that is presented in Figure 3.

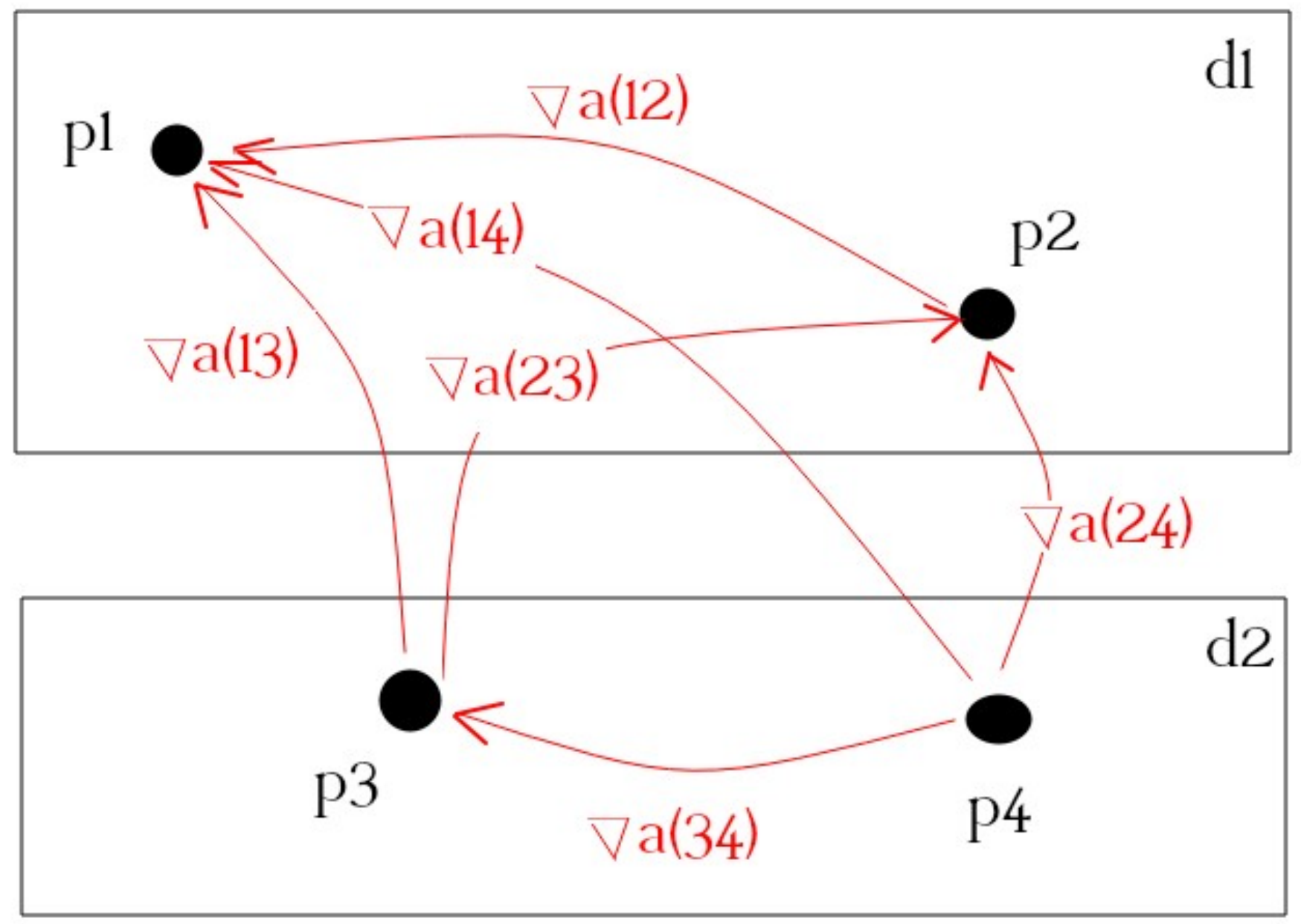

**Figure 3.** An example of the tacit knowledge transfer graph. It represents a situation of two domains ($d_1$ and $d_2$) with the employees $\{p_1, p_2\}$ in $d_1$, and $\{p_3, p_4\}$ in $d_2$. When gradient vanished the arrows are omitted for clarity.

TKTG in Figure 3 is represented by the following adjacency matrix

$$M = \begin{bmatrix} 0 & \nabla a_{12} & \nabla a_{13} & \nabla a_{14} \\ 0 & 0 & \nabla a_{23} & \nabla a_{24} \\ 0 & 0 & 0 & \nabla a_{34} \\ 0 & 0 & 0 & 0 \end{bmatrix}.$$

TKTG is a directed graph G = (P, E), where P is the set of vertices (that are employees) and E is the set of edges, i.e., pairs ($p_i$, $p_j$), weighted by the gradient value $\nabla a_{ji}$. This graph can be represented by the adjacency matrix M, where $M_{ij} = \nabla a_{ji}$. Moreover, we can associate with employee $p_i$ the vector $e_i$ which has 1 at i-th place and all other entries are zeros, which defines an embedding of the employees into the Euclidean space $R^n$, where n=#P (number of employees). We will use below this natural embedding $i$: P → $R^n$ identifying $p_j$ with $e_j$.

We now explore other possibilities to define the gradient of knowledge between employees and connect them with company policies. To this end, we use the sign function

$$sgn(x)=\begin{cases} 1, if x>0 \\ 0, if x=0 \\ -1\, if x<0 \end{cases}.$$

**Definition 1.**

*We can define the gradient of knowledge in the following cases:*

*1.* ***Case I****: The possibility of transfer occurs when either a source or a sink of knowledge exists:*

$$\nabla a_{ij}=\frac{a_{ij}(1)\, b_{ij}(1)-a_{ji}(2)\, b_{ji}(2)}{2}.$$

*2.* ***Case II****: Knowledge gradient vanishes when there is no sink:*

$$\nabla a_{ij}=\frac{sgn\left(a_{ij}(2)\right) a_{ij}(1)\, b_{ij}(1)-a_{ji}(2)\, b_{ji}(2)}{2}.$$

*3.* ***Case III****: Knowledge gradient vanishes when there is no source:*

$$\nabla a_{ij}=\frac{a_{ij}(1)\, b_{ij}(1)-sgn\left(a_{ij}(1)\right) a_{ji}(2)\, b_{ji}(2)}{2}.$$

*4.* ***Case IV****: Knowledge gradient vanishes when there is no source and sink:*

$$\nabla a_{ij} = \frac{sgn\left(a_{ij}(2)\right) a_{ij}(1) b_{ij}(1) - sgn\left(a_{ij}(1)\right) a_{ji}(2) b_{ji}(2)}{2}.$$

For example, if we assume that the gradient must vanish when there is no interest in gaining new knowledge offered by $p_j$ by $p_i$, i.e., there is no enforcement of $p_i$ to use offer of $p_j$ then the Case 2 above is preferable. Then $a_{ij}(2)=0$, so sgn( $a_{ij}(2)$)=0, and so the gradient in this case vanishes as required by the policy.

The gradients, describing the flow of the knowledge (as in the heat flow) and defining TKTG are sufficient for the purposes of this paper. Detailed connection/analogy to the heat transfer/diffusion is described in Appendix C. We also prove here (Theorem C.1) that the transfer of knowledge suggested by TKTG leads to 'equalizing' knowledge deficiency among employees.

### 4.2.1 Simplified case

We also propose a simplified model that can be implemented in companies where there is no possibility of organizing extensive questionnaires. Since questioning of all employees about all other employees (ar even groups of specific discipline) is of quadratic complexity in the number of employees, $O((\#P)^2)$, therefore, this optimization allows to adjust data gathering to the company resources.

In this approach, each employee $p_i$ is asked if there is the demand to learn from other employee, and if so, which one. Assume that such source is the employee $p_j$, then we set $\nabla a_{ij}=1$, and we set it to zero otherwise. Call it **the Minimal Case** for future reference. Other possibilities of definition of the gradient can be made, e.g., we can define the knowledge gradient as the maximum or minimum of the answers of the source and sink. Then the adjacency matrix M consists only values 0 and 1 and can be related to the cases from the Definition 1 as a rounding of the adjacency matrix defined using the gradients from the Definition 1.

Such simplified adjacency matrix can be easily constructed, however, it contains less precise information about the real need for knowledge transfer.

In order to save even more resources during survey, a manager can designate a teacher from each domain, and then during the meeting with teams ask who needs knowledge from which domain and construct TKTG and its adjacency matrix.

### 4.3. TKTG properties

From the definition of the adjacency matrix for the knowledge transfer graph one can deduce the following important properties:

**Property 1**

*If* $M_{ij}=\nabla a_{ij}=0$ *for all j (i-th row vanishes) then it means that (using cases from Definition 1):*

*1. Case I:* $p_i$ *do not want to learn from everyone AND does not want to be taught by anyone.*

*2. Case II: Case I OR* $p_i$ *do not want to learn.*

*3. Case III: Case I OR* $p_i$ *do not want to be taught by anyone.*

*4. Case IV:* $p_i$ *do not want to learn from everyone OR do not want to teach by everyone.*

*5. Minimal Case:* $p_i$ *do not want or need to learn in general.*

These cases are diagnostic information that allows for identification of potential problems in career development of a given employee, or potential interpersonal conflicts.

Interpreting this case in TKTG we will see the point $p_i$ with no entering arrows.

Likewise we can focus on the vanishing column:

**Property 2**

*If* $M_{ij}=\nabla a_{ij}=0$ *for all i (j-th column vanishes) then it means that (using cases from Definition 1):*

1. *Case I:* $p_j$ *do not want to learn from everyone AND do not want to be taught by everyone.*

2. *Case II: Case I OR nobody wants to learn from* $p_j$*.*

3. *Case III: Case I OR* $p_j$ *do not want to teach anyone.*

4. *Case IV:* $p_j$ *do not want to learn from everyone OR do not want to teach everyone.*

5. *Minimal Case:* $p_j$ *do not represent the knowledge needed by anyone.*

On TKTG the point $p_j$ in this case has no outgoing arrows.

Both these situations can be identified in the diagonalized form of the adjacency matrix since the correspond to the eigenvalue 0 and its eigenvector.

### 4.4 Dynamics of TKTG

The adjacency matrix M of TKTG defines dynamics of knowledge transfer via the following equation

$$M\,p_i=\sum c_j\,p_j\,,\,where\,c_j\neq 0,$$

where the embedding $p_i \leftrightarrow e_i$ is used[3]. It can be represented on TKTG starting from $p_i$ and moving along the edges emanating from this point to the next vertices. Therefore, looking on the right hand side of the equation one can identify all employees that can be taught by $p_i$. That defines discrete dynamics on TKTG, where each step of evolution results in applying M on the selected vector v composed with vertices

$$v_{n+1} = M v_n, n \geq 0,$$

where $v_0$ is some initial vector. The natural question is to ask if one can select initial vector that simplifies the dynamics of knowledge transfer. The one possible solution is to use Jordan decomposition of M that will be the topic of the next subsection.

### 4.4.1 Jordan decomposition and optimal learning groups

Since the matrix M is, in general, not symmetric, therefore the more suitable choice is the Jordan decomposition than the eigenvalue decomposition.

As a result of the decomposition the matrix M is decomposed into diagonal blocks depending on the type of eigenvalues. We find the matrix A that gives J such that $J = A^{-1} MA$ is a matrix in Jordan normal form. We consider the possible types of blocks and its result on the dynamics determined by M.

**One-dimensional Jordan block [λ]:** The eigenvector v corresponding to this eigenvalue λ defines the fixed point of the knowledge transfer dynamics given by

$$Mv = \lambda v.$$

[3]Note that in the sum we do not include j=i case since the diagonal of M consists of zeros. This corresponds to the existence of no loops starting and ending in $p_i$.

If $v=\sum c_i p_i$ for non-vanishing coefficients $c_i$ then the dynamics mixes the employees that enters into v and no other employees. We can trace exactly which employees teach others by $M p_i=\sum f_j p_j, where f_j \neq 0$. We therefore have the following

**Corollary 1**

*An eigenvector can be used to arrange knowledge exchange sessions scheduled closely together, e.g., in the same day, since it describes all the employees that are required for teaching sessions.*

**Two-dimensional Jordan block** $\begin{bmatrix} \lambda & 1 \\ 0 & \lambda \end{bmatrix}$: One eigenvector $v_1$ and one generalized eigenvector $v_2$ are defined as follows

$$Mv_1=\lambda v_1,$$

$$Mv_2=\lambda v_2+v_1.$$

This means that

$$(M-\lambda I)v_2=v_1,$$

$$(M-\lambda I)^2 v_2=0.$$

The first equation shows that the matrix $M-\lambda I$ describe the knowledge transfer dynamics that transforms $v_2$ components into $v_1$ components. We therefore have dynamics that transforms $v_1 \rightarrow v_1$, and $v_2 \rightarrow v_1$. We therefore infer

**Corollary 2**

For a two-dimensional Jordan block, we have dynamics between two eigenvectors. The components of these two vectors determine the group of employees that can participate in the knowledge exchange sessions, grouped closely together, e.g., on the same day.

**Higher-dimensional Jordan blocks**: For n-dimensional block this part describes cyclic vectors $v_1,\ldots, v_n$, that fulfills

$$\begin{gathered}(M-\lambda I)v_n=v_{n-1}\\ \ldots\\ (M-\lambda I)v_1=0\end{gathered}.$$

Then, by analogy to the previous cases, the components of these eigenvectors can be used to construct the closed groups for arranging knowledge transfer sessions optimally, one after another.

Eigenvectors for specific eigenvalues define invariant subspaces with respect to the dynamics defined by M. Therefore, the employees involved in the components of the eigenvectors are preserved under the dynamics defined by M. However, examining TKTG dynamics from two eigenvector components reveals that their contributions can cancel, and therefore, the employee they should teach does not have to appear in the eigenvector or cyclic vector. This is because eigenvectors have real coefficients, and we do not attach any meaning to them since we treat these vectors only as information about the invariant subspace spanned by the employees. To track all trained employees, it is therefore essential to examine the dynamics of each employee within the invariant subspace under M. As a result, we can form an algorithm for planning knowledge transfer sessions:

**Algorithm 1**

*For planning of knowledge sharing sessions do the following steps:*

*1. For each eigenspace of Jordan decomposition extract employees entering as a components of the eigenvectors from this space – employee $p_i$ is a teacher when the i-th entry of some eigenvector from this space is nonzero. This gives the list of teachers for a given eigenspace $P_i$.*

*2. Trace the dynamics of each employee from the previous step by applying M and note the trained employees, i.e., compute $Mp_k$ from $p_k \in P_i$.*

*3. Arrange schedule where results from analyzing an eigenspace are close to each other, e.g., in the same day.*

Theorem C.1 from Appendix C states that this Algorithm, in fact, leads to 'smoothing' knowledge deficiencies among employees.

From the preceding discussion we have that in the second step we obtain trained employees from $P_i$, and additional ones that do not enter as components of eigenvectors of an eigenspace due to cancellation of coefficients of the component.

The general algorithms for Jordan normal form decomposition have in general time complexity of order $O(n^3)$, where n is the size of square matrix (the number of employees under consideration), see, e.g., (Kågström, Ruhe, 1980), (Zeng, Li, 2021). However, the problems of scheduling of classes (UCTP - University Course Timetable Problem) are NP-complex problems in general (Mühlenthaler, 2015). The preparatory step that is adaptation of the algorithm presented above can lead as a heuristic for speed up computations of UCTP algorithms.

Jordan decomposition is one of the many methods of bringing a matrix to a normal form, e.g., Schur decomposition. We use it here since it is straightforward to interpret in terms of knowledge transfer. However, more numerically stable methods can also be used, while maintaining explainability. For example, the Schur decomposition brings a matrix to its upper triangular form, which is more numerically stable; however, the eigenspaces are larger since they typically contain more cyclic vectors. That will result in a larger closed set of employees for whom the learning sessions can be scheduled on the same day on one side. On the other side, the results will depend less on the inaccuracies in constructing the adjacency matrix for TKTG from questionnaire data. This issue can be mitigated by using the Minimal Case or applying the signum function to each entry of the adjacency matrix.

We can also obtain information about the importance of individual employees in the knowledge transfer process described by M. We describe it in the next part.

#### 4.4.2 Differentiating the employees needs in knowledge transfer process

To define the employees that have greater knowledge demand we must use a measure that counts the number of inward edges to this nodes. We present three approaches for the adjacency matrix M.

**Definition 2. [Counting inward edges]**

*Denote by sgn(M) the matrix with element ij being sgn($M_{ij}$). Since $M \geq 0$, so the entries of sgn(M) are either 0 or 1.We can now construct the score s($p_i$) of the employee i by*

$$s(p_i)=\frac{1}{n}\sum [sgn(M)]_{ik}.$$

This is a quantity that measures the number of inward links to the total number of nodes, that can potentially have inward edges. The employees with higher score have higher knowledge transfer demands.

For M that is primitive, see Appendix A, we have an eigenvector v associated to the highest eigenvalue, which has all positive entries. Normalizing it, |v|=1, we get so called eigenvector centrality: the i-th entry of normalized v measures the weighted inward edges pointing into $p_i$. As before, the higher eigenvector centrality of a vertex, the higher knowledge demand of the employee $p_i$. We can extract this positive vector associated to the maximal eigenvalue using iterative procedure applied to the matrix $\frac{1}{\lambda_{max}} M$ according to the Theorem A.2 of Appendix A.

The third score, which can be used to identify employees with high knowledge demands for M that are not necessarily primitive, is a weighted PageRank algorithm (see Appendix B). We can also apply the standard PageRank algorithm to the sgn(M) matrix. The input is the adjacency matrix M, and as a result, we get a vector of scores s, where $s_i$ is the score of $p_i$, which measures the scoring of the neighbors weighted by outward scores as described in Appendix B.

We will provide examples in the next section.

## 5. An example applications

In this section we provide some examples that explain the model and illustrate it on some specific examples that identifies some general phenomena we start from small learning groups and extend the results to high number of group members.

### 5.1. Example 1

We imagine the small construction site, where two collaborating expert groups:

- $d_1$: A group of two electricians, $p_1$, and $p_2$.

- $d_2$: A single plumber, $p_3$.

The specialists are:

- $p_1$: An experienced electrician who requires significant plumbing knowledge ($a_{13}(2)=-1$) to improve their performance, and (s)he can learn from plumber without any personal obstacles ($b_{13}(2)=1$). (S)he also can transfer the experience to the junior $p_2$ without any personal problems ($a_{21}(1)=1$, $b_{21}(1)=1$).

- $p_2$: A junior electrician who needs foundational knowledge from $p_1$ ($a_{21}(2)=-1$), and have no obstacles to work with him/her ($b_{13}(2)=1$). Obviously, (s)he cannot learn anything from $p_1$ ($a_{12}(1)=0$, $b_{12}(1)=1$), and $p_3$ ($a_{32}(1)=0$, $b_{23}(1)=1$). The b's are set to 1 since $p_2$ is open for professional contacts.

- $p_3$: An experienced plumber who can share basic knowledge with $p_1$ ($a_{13}(1)=1$, $b_{13}(1)=1$) but is reluctant to interact with the inexperienced electrician $p_2$, which yields ($a_{23}(1)=1$, $b_{23}(1)=0$).

Assume that after questionnaires we infer the **A** and **B** matrices along with the gradients presented in Table 2.

| | **$p_1$** | **$p_2$** | **$p_3$** |
|---|---|---|---|
| **$p_1$** | 0 | [0, 0, 1, 1]<br>$\nabla a_{12}=0$ | [1, -1, 1, 1]<br>$\nabla a_{13}=\frac{1-(-1)}{2}=1$ |
| **$p_2$** | [1, -1, 1,1] | 0 | [1, 0, 0, 1] |

|  | $\nabla a_{13} = \frac{1-(-1)}{2} = 1$ |  | $\nabla a_{23} = 0$ |
|---|---|---|---|
| **p₃** | [0, 0, 1, 1]<br><br>$\nabla a_{31} = 0$ | [0, 0, 1, 0]<br><br>$\nabla a_{32} = 0$ | 0 |

**Table 2**. Presents the data infer from questionnaires among specialists at workplace.

This generates knowledge graph from Figure 4.

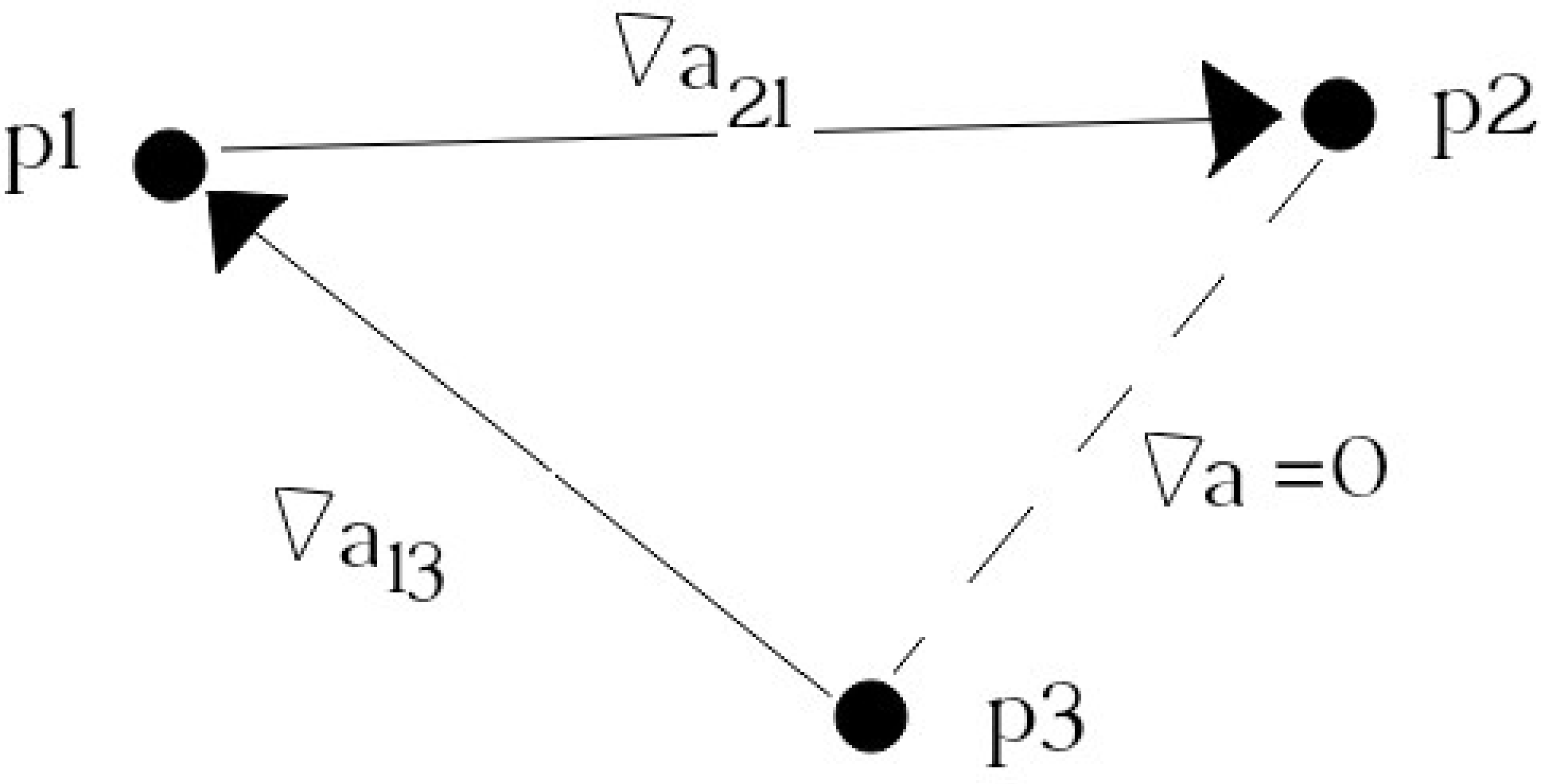

**Figure 4.** Knowledge transfer graph.

Diagonalizing the adjacency matrix M one gets the following Jordan form

$$J = \begin{bmatrix} 0\,1\,0 \\ 0\,0\,1 \\ 0\,0\,0 \end{bmatrix},$$

with the following (cyclic) eigenvectors

$$v_1 = \begin{bmatrix} 0 \\ 0 \\ 1 \end{bmatrix} = e_3, v_2 = \begin{bmatrix} 1 \\ 0 \\ 0 \end{bmatrix} = e_1, v_3 = \begin{bmatrix} 0 \\ 1 \\ 0 \end{bmatrix} = e_2,$$

that is $v_1$ is associated with $p_3 = e_3$, $v_2$ is associated with $p_1=e_1$, and $v_3$ vector is associated with $p_2=e_2$ . Therefore eigenvectors in this simple case determine not groups but single employees – each vector has only a single non-zero component. The J matrix determines now the knowledge transfer direction:

$$v_1 \overset{J}{\rightarrow} v_2 \overset{J}{\rightarrow} v_3,$$

that corresponds to

$$e_3 \overset{M}{\rightarrow} e_1 \overset{M}{\rightarrow} e_2,$$

which makes the following transfer of knowledge in learning group[4]:

$$p_3 \overset{M}{\rightarrow} p_1 \overset{M}{\rightarrow} p_2.$$

Since all three vectors $v_i$ belongs to the same eigenspace, so for effectiveness, the classes must be arranged in the same day one after another.

The example is trivial since the (singleton) groups can be read off from the graph in Figure 3. The groups consists of a single employee, which is reflected in the zero eigenvalue in the Jordan decomposition in J matrix. The map described by J matrix reflects intuitive thinking that the knowledge transfer should be as follows: the plumber $p_3$ teaches the experienced electrician $p_1$, and experienced electrician $p_1$ teaches $p_2$. In this simple example the Jordan decomposition only confirms what is visible from the graph in Figure 3.

[4]These learning groups described by generalized eigenvectors are degenerate in the sense that each eigenvector describes a single employee. In general, eigenvectors can mix employees describing a group that form the group.

### 5.2 Example 2

We focus on three employees P={$p_1$, $p_2$, $p_3$}, with the following adjacency matrix M for TKTG:

$$M=\begin{bmatrix}0&1&0\\0&0&1\\0&1&0\end{bmatrix}.$$

One can observe that $p_1$ do not teach, $p_2$ transfer knowledge to $p_1$ and $p_3$, and $p_3$ transfer knowledge to $p_2$. Bringing the the Jordan form one gets

$$J=\begin{bmatrix}-1&0&0\\0&0&0\\0&0&1\end{bmatrix},$$

with the corresponding eigenvectors:

$$v_1=\begin{bmatrix}1\\-1\\1\end{bmatrix}=e_1-e_2+e_3,\ v_2=\begin{bmatrix}1\\0\\0\end{bmatrix}=e_1,\ v_3=\begin{bmatrix}1\\1\\1\end{bmatrix}=e_1+e_2+e_3.$$

We consider independently each eigenspace:

1. $\mathbf{v_1}$**:** consists of all employees. Tracing the dynamics, we have: ($p_1$ do not teach), ($p_2$ teach $p_1$ and $p_3$) and ($p_3$ teaches $p_2$)

2. $\mathbf{v_2}$**:** is eigenvalue zero vector so component $p_1$ teach no one;

3. $\mathbf{v_3}$**:** again consist all employees and it results in the same groups as the case 1.

### 5.3 Example 3

Let us take the group of three employees that all require to learn from each other. The adjacency matrix is of the form

$$M = \begin{bmatrix} 0 1 1 \\ 1 0 1 \\ 1 1 0 \end{bmatrix},$$

that is all employees are both teachers and learners. The Jordan decomposition gives

$$J = \begin{bmatrix} -1 & 0 & 0 \\ 0 & -1 & 0 \\ 0 & 0 & 2 \end{bmatrix},$$

and eigenvectors

$$v_1 = \begin{bmatrix} -1 \\ 0 \\ 1 \end{bmatrix} = -e_1 + e_3, v_2 = \begin{bmatrix} -1 \\ 1 \\ 0 \end{bmatrix} = -e_1 + e_2, v_3 = \begin{bmatrix} 1 \\ 1 \\ 1 \end{bmatrix} = e_1 + e_2 + e_3.$$

We have three eigenspaces to analyze:

1. $\mathbf{v_1}$**:** ($p_1$ teaches $p_2$ and $p_3$) and ($p_3$ teaches $p_1$ and $p_2$),

2. $\mathbf{v_2}$**:** ($p_1$ teaches $p_2$ and $p_3$) and ($p_2$ teaches $p_1$ and $p_3$),

3. $\mathbf{v_3}$**:** ($p_1$ teaches $p_2$ and $p_3$) and ($p_2$ teaches $p_1$ and $p_3$) and ($p_3$ teaches $p_1$ and $p_2$).

It gives us either option 1 and 2 for organizing learning sessions in two different days, or the option 3 that suggest to organize all three sessions in a single day.

### 5.4 Example 4

As a final example, we provide example of a Monte Carlo simulated example with 10 employees. The randomly generated adjacency matrix is presented in Figure 5 and 6.

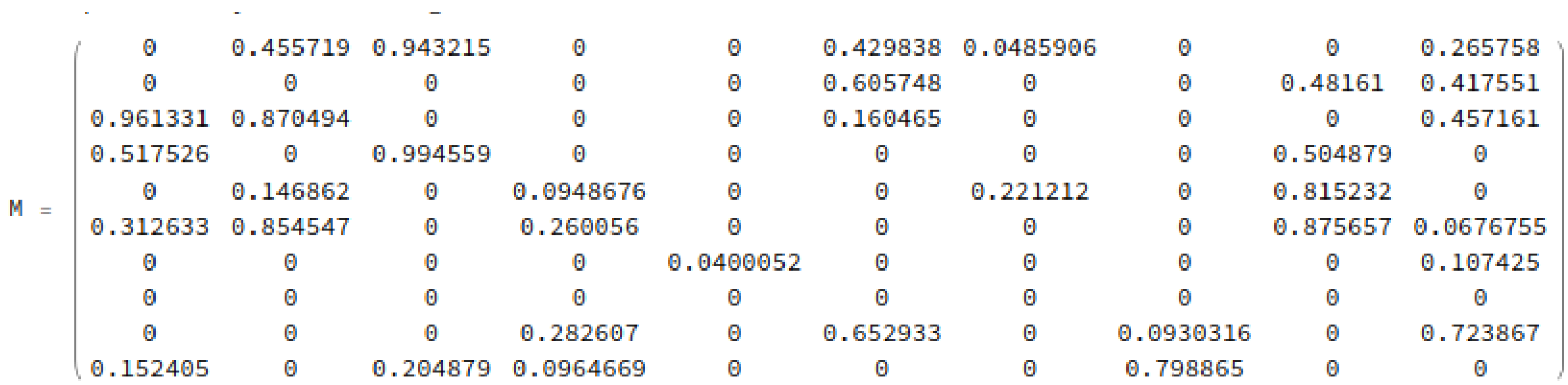

$$M = \begin{pmatrix} 0 & 0.455719 & 0.943215 & 0 & 0 & 0.429838 & 0.0485906 & 0 & 0 & 0.265758 \\ 0 & 0 & 0 & 0 & 0 & 0.605748 & 0 & 0 & 0.48161 & 0.417551 \\ 0.961331 & 0.870494 & 0 & 0 & 0 & 0.160465 & 0 & 0 & 0 & 0.457161 \\ 0.517526 & 0 & 0.994559 & 0 & 0 & 0 & 0 & 0 & 0.504879 & 0 \\ 0 & 0.146862 & 0 & 0.0948676 & 0 & 0 & 0.221212 & 0 & 0.815232 & 0 \\ 0.312633 & 0.854547 & 0 & 0.260056 & 0 & 0 & 0 & 0 & 0.875657 & 0.0676755 \\ 0 & 0 & 0 & 0 & 0.0400052 & 0 & 0 & 0 & 0 & 0.107425 \\ 0 & 0 & 0 & 0 & 0 & 0 & 0 & 0 & 0 & 0 \\ 0 & 0 & 0 & 0.282607 & 0 & 0.652933 & 0 & 0.0930316 & 0 & 0.723867 \\ 0.152405 & 0 & 0.204879 & 0.0964669 & 0 & 0 & 0 & 0.798865 & 0 & 0 \end{pmatrix}$$

**Figure 5.** Randomly generated adjacency matrix.

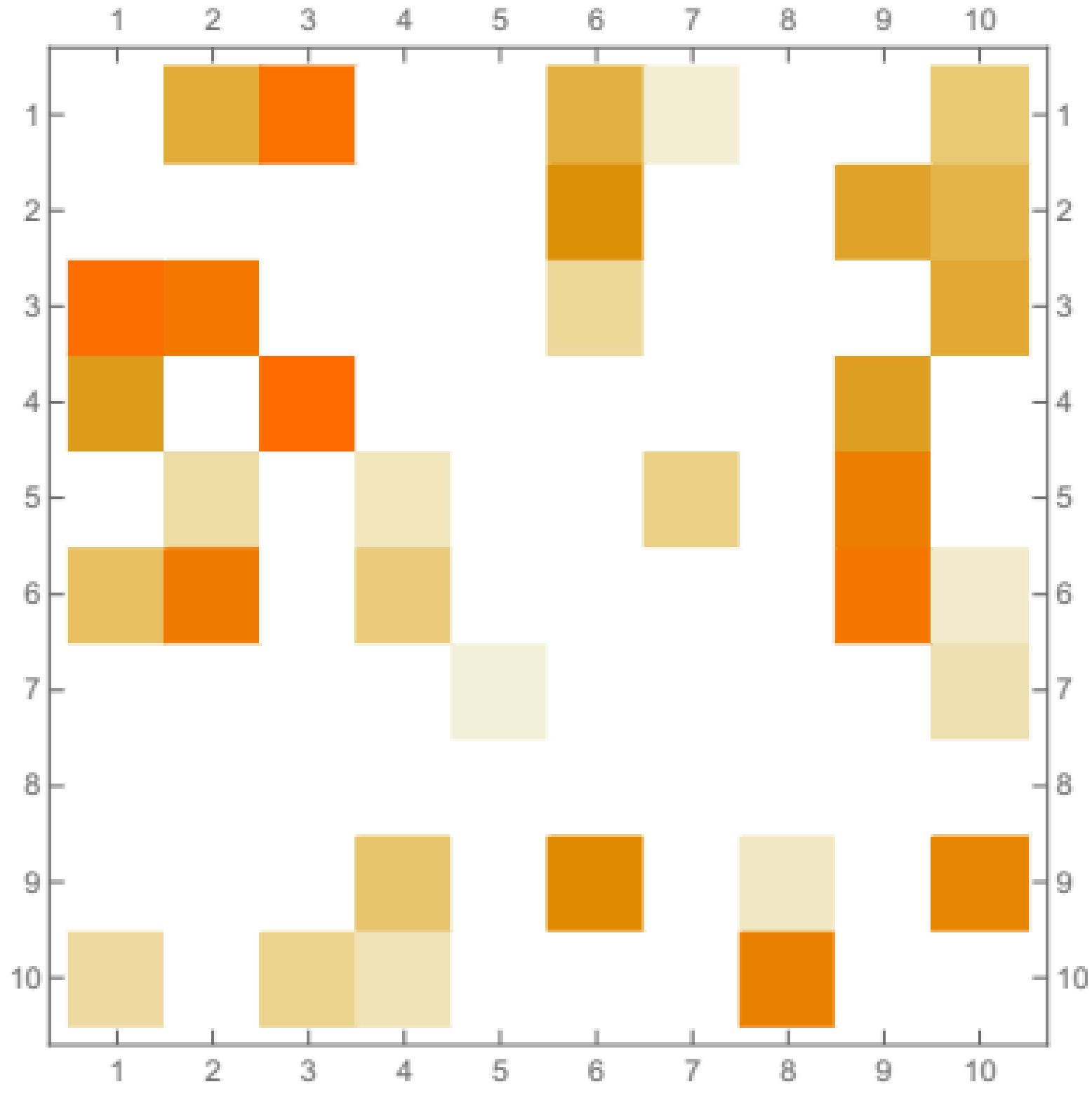

**Figure 6.** Adjacency matrix – graphical representation.

As a result of Jordan decomposition we obtain 10 eigenvalues defining one-dimensional eigenspaces. Analyzing dynamics of adjacency matrix one obtains the same 9 result not containing $p_8$ as a teacher for each eigenspace, and one case where each employee is a teacher, which is presented in Table 3.

| Teacher | Learning group |
|---|---|
| $p_1$ | $p_3$, $p_4$, $p_6$, $p_{10}$ |
| $p_2$ | $p_1$, $p_3$, $p_5$, $p_6$ |
| $p_3$ | $p_1$, $p_4$, $p_{10}$ |
| $p_4$ | $p_5$, $p_6$, $p_9$, $p_{10}$ |
| $p_5$ | $p_7$ |
| $p_6$ | $p_1$, $p_2$, $p_3$, $p_9$ |
| $p_7$ | $p_1$, $p_5$ |
| $p_8$ | $p_9$, $p_{10}$ |
| $p_9$ | $p_2$, $p_4$, $p_5$, $p_6$ |
| $p_{10}$ | $p_1$, $p_2$, $p_3$, $p_6$, $p_7$, $p_9$ |

**Table 3.** Content of learning groups. The group where $p_8$ is a teacher is not represented in every eigenspace.

### 5.5 Example 5 – Monte Carlo experiments

As a next example we check the averaged dimension of eigenspaces with respect to the percentage of zeros in the adjacency matrices. We use Monte Carlo method to simulate results using the algorithm presented in Figure 7.

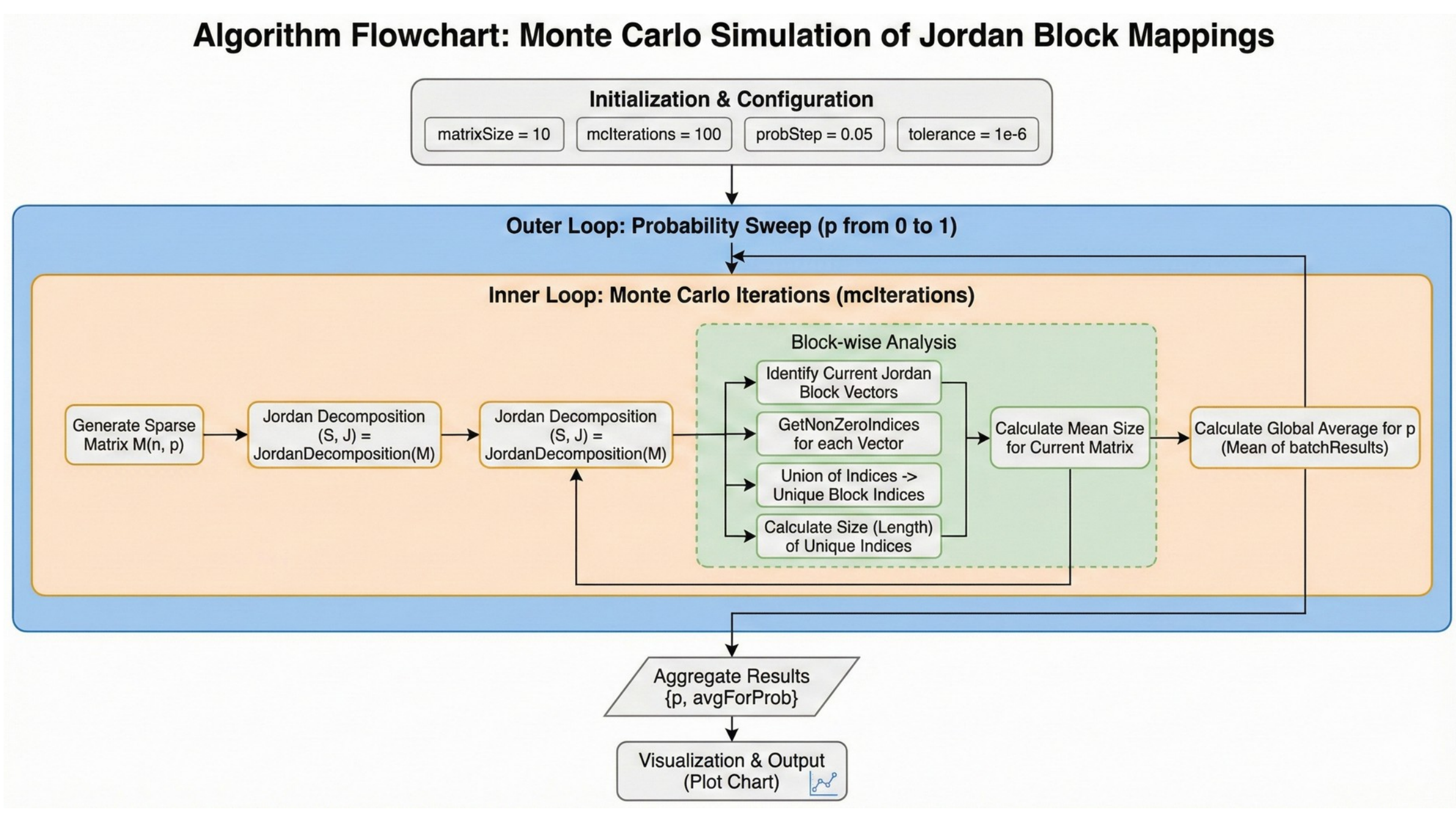

**Figure 7:** Monte Carlo algorithm for calculating average groups depending on the percentage number of zeros [figure generated using Gemini AI based on the code in Wolfram Mathematica/Wolfram Language].

We selected adjacency matrix dimensions n=10, 20, 30, 40. The results are presented in Figure 8.

**Figure 8.** The averaged size of set of unique base vectors for each eigenspace (OY axis) with respect to the zeroProbablity (OX axis). Each point represents 100 Monte Carlo runs.

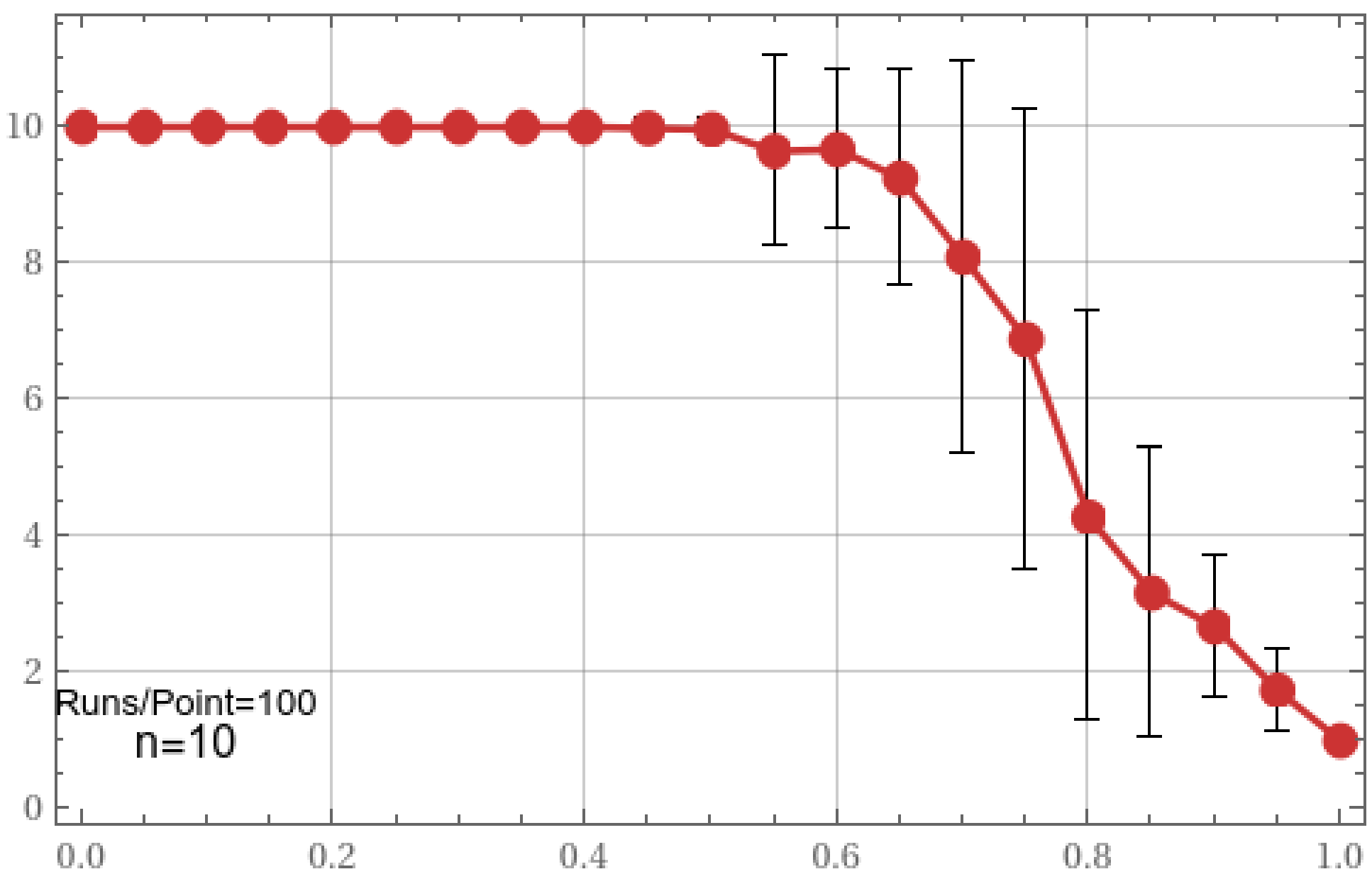

a) n=10

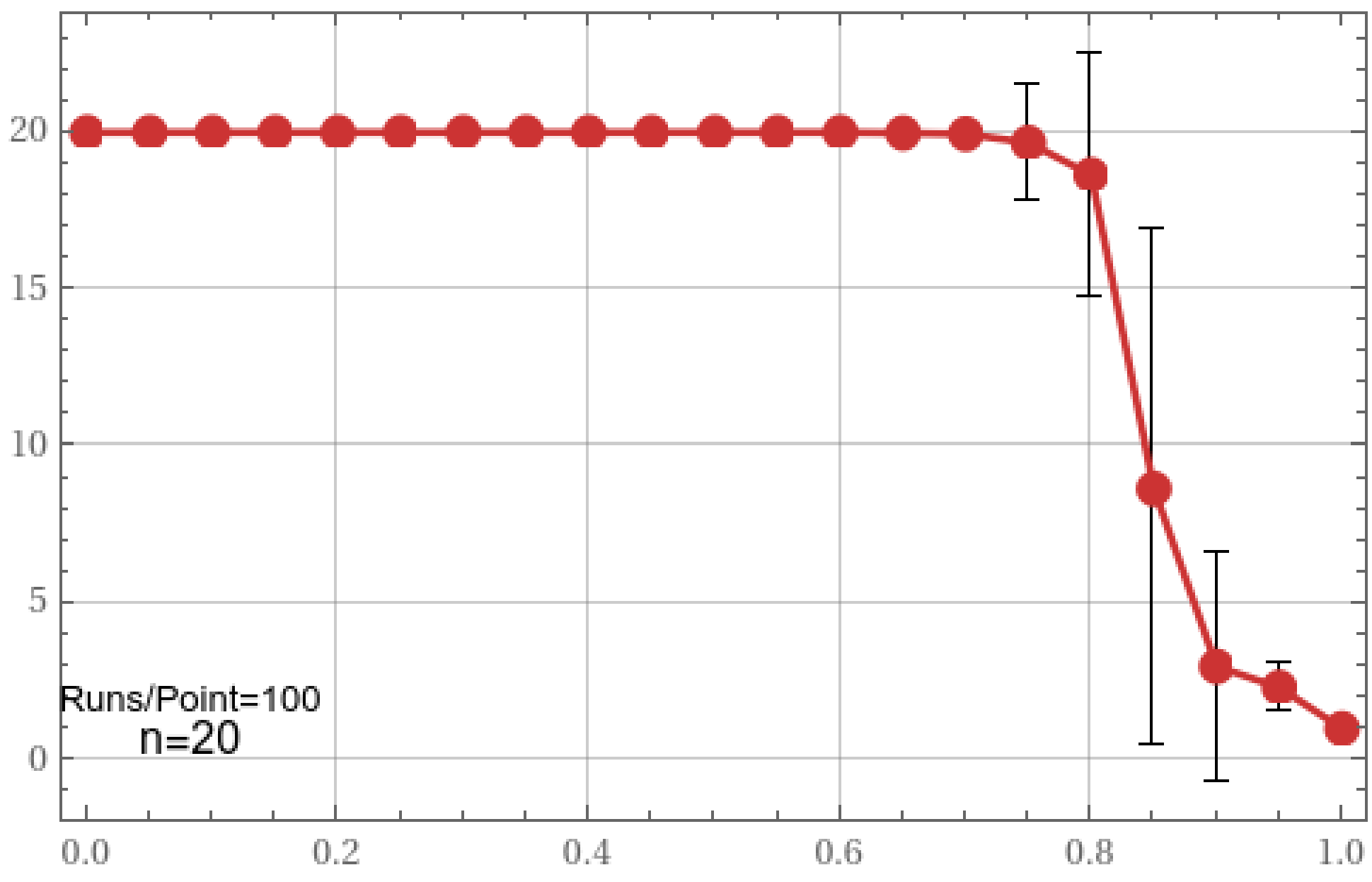
20
15
10
5
0
0.0
0.2
0.4
0.6
0.8
1.0
Runs/Point=100
n=20

b) n=20

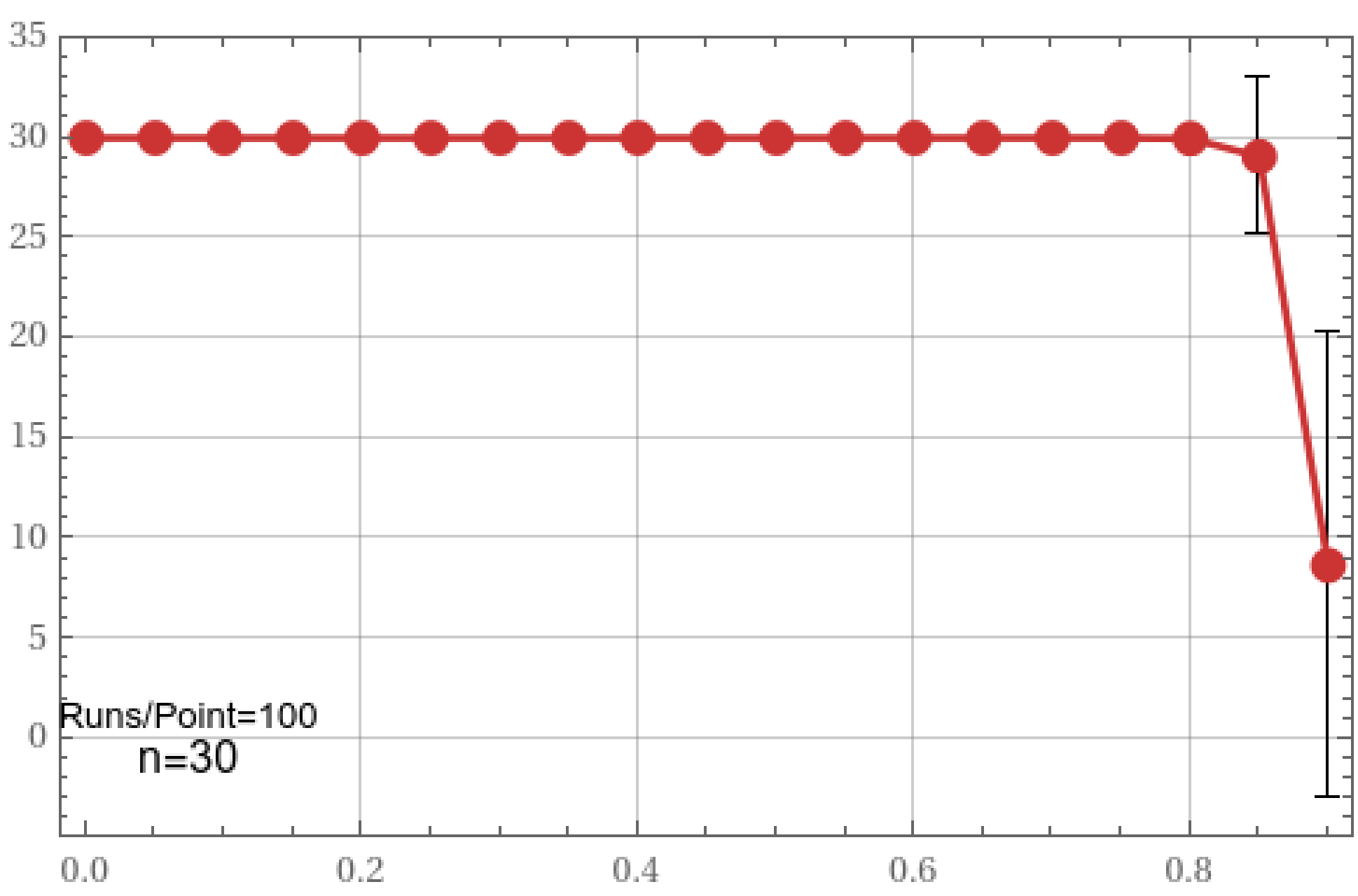
35
30
25
20
15
10
5
0
0.0
0.2
0.4
0.6
0.8
Runs/Point=100
n=30

c) n=30

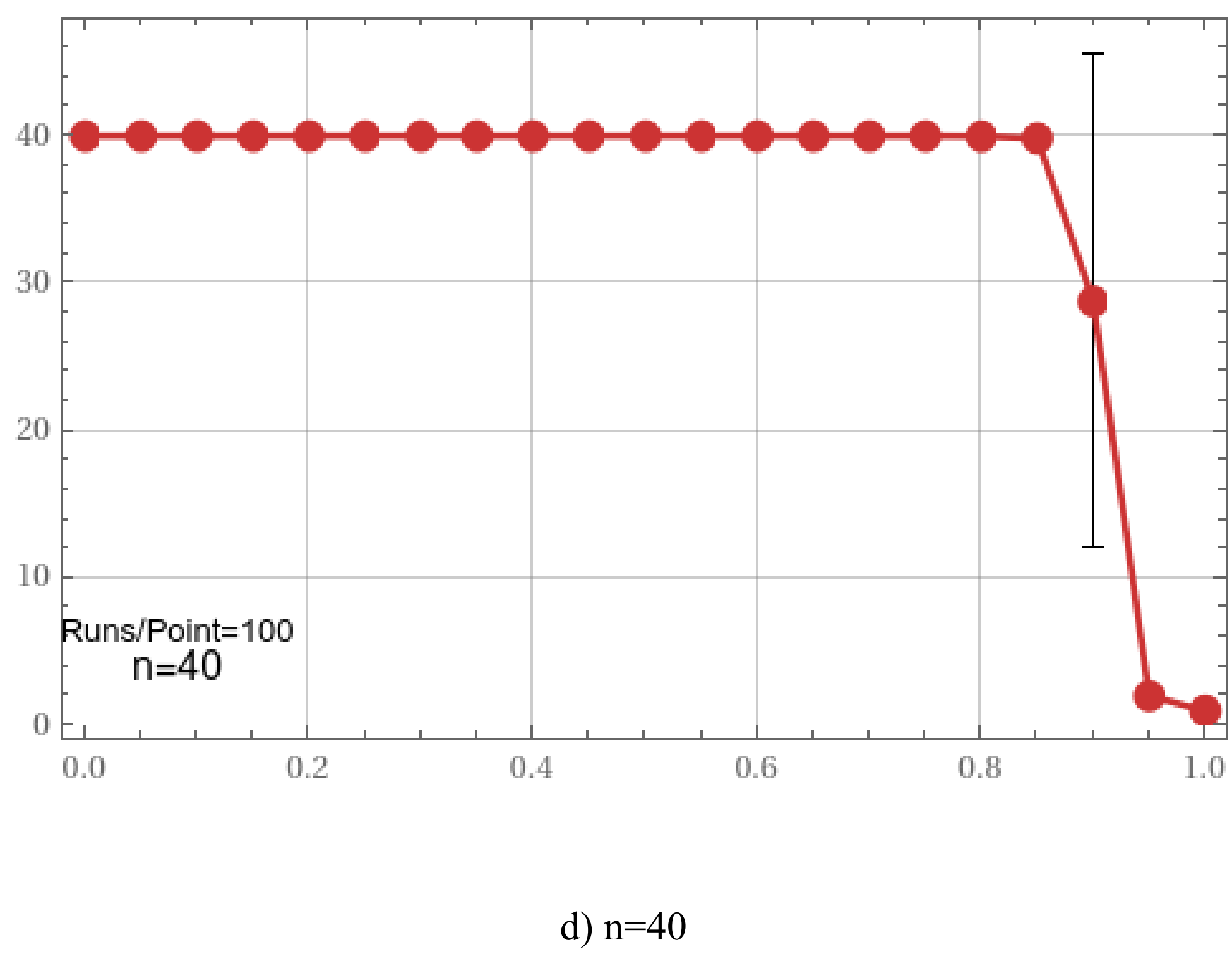

d) n=40

One can observe that the deflection point for a large percentage of zero elements in the adjacency matrix shifts with the size of the adjacency matrix (i.e., the number of employees). This resembles a phase transition observed in physics when analyzing Critical Phenomena, as described in Binney, Dowrick et al. (1992). It is observed that when the matrix has few non-zero off-diagonal entries, the eigenspaces are spanned by vectors consisting of only a few base vectors. That suggests an obvious:

**Conclusion 1.**

*When knowledge demands are low (measured as a number of off-diagonal zeros in adjacency matrix of TKTG) then the average size of optimally planned classes in the same time period (say, day) is less, and the schedule can be more flexible.*

In the next section, we present the application of the algorithm to school-related applications as a other possible application of our method outside of the business.

**6. Application to planning classes in educational institutions**

The presented ideas, slightly adjusted, can be also used in schools and universities. In this case we have two types of actors:

- teachers: $T=\{t_1\ldots, t_N\}$,
- pupils/students collected into groups: $G=\{g_1,\ldots, g_K\}$.

Since only the groups require learning from teachers, we can use Minimal Case to construct adjacency matrix setting 1 in the entry representing the fact that teacher $t_i$ teaches the group $g_j$. The matrix has the block form

$$M=\left[\begin{array}{c|c} 0 & A \\ 0 & 0 \end{array}\right],$$

where the block A of dimension NxK, represents the teaching of groups by teachers.

As a simple example consider two groups $G=\{g_1, g_2\}$ and two teachers $T=\{t_1, t_2\}$. Assume that $t_1$ teaches $g_1$ and $g_2$, and $t_2$ teaches only $g_2$. Using Minimal Case for constructing gradients, and arranging the columns/rows of the adjacency matrix in the order: $g_1$, $g_2$, $t_1$, $t_2$ we get

$$M=\begin{bmatrix}0&0&1&0\\0&0&1&1\\0&0&0&0\\0&0&0&0\end{bmatrix}.$$

The Jordan decomposition gives

$$J=\begin{bmatrix}0&1&0&0\\0&0&0&0\\0&0&0&1\\0&0&0&0\end{bmatrix},$$

with eigenvectors

$$v_1=\begin{bmatrix}0\\1\\0\\0\end{bmatrix}=g_2, v_2=\begin{bmatrix}0\\0\\0\\1\end{bmatrix}=t_2, v_3=\begin{bmatrix}1\\0\\0\\0\end{bmatrix}=g_1, v_4=\begin{bmatrix}0\\0\\1\\-1\end{bmatrix}=t_1-t_2.$$

We have the following dynamics given by cyclic vectors with eigenvalues 0: $v_2 \rightarrow v_1$, and $v_4 \rightarrow v_3$. Noting the dynamics of the components of input vectors by the dynamics given by M we can express the splitting the teaching into two sessions:

1. $v_2 \rightarrow v_1$, gives: ($t_2$ teaches $g_2$),

2. $v_4 \rightarrow v_3$, gives: ($t_1$ teaches $g_1$), ($t_1$ teaches $g_2$), ($t_2$ teaches $g_2$).

The last case represents, say the day, where all classes happens. This is optimal in the sense that all teachers and all groups are present.

**7. Limitations and Future Research**

As the present study focuses on the formal development of a theoretical framework, empirical validation and organization-specific implementations are intentionally left for future research.

The model is a general framework - a metamodel -for assessing knowledge transfer. Each specific implementation requires adjustment in several areas:

1. **Questionnaire Design:** A questionnaire must be carefully designed to reflect the organizational structure, culture, and domains, while also accounting for psychological biases. It should also take into account possible insincere answers to questions, e.g., Attention Checks, Reverse Coding, Projective Techniques (Indirect Questioning), Face-Saving Introductions, or Behavioral Frequency.

2. **Weighting:** The construction of the values $a_{ij}$ can, in principle, include weights ($v_i$,$v_j$) representing the cost of knowledge transfer activities, such as organizing meetings. In the simplest case, these can be set to 1.

3. **Dynamic Model:** While we have described a static model, a dynamic model would consist of this static model as a series of time snapshots. Implementation of a dynamic model requires conducting surveys on a regular basis, which increases cost but also the level of detail on knowledge transfer in the organization.

**8. Implications**

The proposed method has implications for various aspects of knowledge management:

- **Knowledge Management Practices:** The method helps to identify key employees and groups for knowledge sharing, thereby improving the organization of the knowledge transfer process.
- **Economic Activities:** While knowledge transfer incurs a short-term cost in terms of employee time, it ultimately aims to create more effective teams in the long run. This supports well-organized knowledge sharing and helps build a company founded on knowledge and expertise.

- **Policy Making:** Knowledge sharing is a crucial element of a well-organized company, and its aspects should be integrated when developing corporate and employee policies. The proposed method helps formalize the knowledge transfer procedure and can be used to create a clear, structured policy for this activity.

- **Social Development:** Unidentified knowledge demands and transfer "blockers" among employees can lead to serious social and psychological problems, including quiet quitting and employee turnover. A well-crafted knowledge transfer policy, supported by this method, helps to develop staff and supports their need for self-development, thus improving the organizational culture.

The presented model influences:

- **Optimal scheduling and operational efficiency** – identifying employees related to each other through a teaching-learning relationship allows managers to expedite the planning of the learning process and pinpoint ‘learning circles’. This also helps reduce redundant interactions and plan the cost of learning efficiently.

- **Talent management –** the questionnaires, especially b coefficients, allow HR to identify potential barriers between employees that prevent knowledge from flowing. Moreover, eigenvectors centrality allows us to identify key players in knowledge flow – “sources” and “sinks”.

- **Risk management in knowledge transfer –** the TKTG allows for the identification of critical players in knowledge transfer (those with the highest outward edges – highly occupied teachers). That helps plan for the risk of losing highly skilled employees. Moreover, the inclusion of external contractors in the analysis helps identify external dependence and evaluate their importance in the project's success.

- **Organization culture and policy** – the model enables planning for knowledge transfer and learning groups' content. Effective planning can increase the morale of employees who are driven by personal development and decrease the rate of (quiet) quitting. The self-learning culture can also help Generation Z employees establish a stronger connection with the organization through learning groups (Nieżurawska, Niemczynowicz, and Kycia, 2023).

## 9. Conclusions

The model allows for the identification of tacit knowledge assets and their transfer among employees in market-oriented companies. The tacit knowledge transfer graph, a key construct of the model, enables monitoring of tacit knowledge, facilitating the construction of a true knowledge-based organization with efficient knowledge flow. Such an approach requires an interdisciplinary approach. In this case, concepts from physics were employed; particularly, Fourier's law of heat conduction was adapted.

Graph theory allows managers to visualize tacit knowledge transfer. Moreover, the need for tacit knowledge can be calculated, as can the willingness to share it. This allows for promoting employees willing to transfer knowledge, rather than those unwilling to do so. Some workers prefer to maintain their unique competencies, which are usually based on tacit knowledge, without disclosing them.

Observation, experience, practice, and critical thinking are essential factors that foster an increase in tacit knowledge. Since tacit knowledge is tied to individuals, it is necessary not only to estimate its presence but also to facilitate its sharing among other employees. The proposed model enables the formation of self-learning, multidisciplinary groups that can boost knowledge sharing.

**Conflict of Interest**

Authors declare no conflict of interests.

**Acknowledgments**

We would like to thanks anonymous Referees, whose kind and constructive comments helped us to improve the paper.

Figure 7 is generated using Gemini AI based on the code in Wolfram Mathematica/Wolfram Language.

RK and AN are acknowledge the support of COST CA24122 action.

RK and AN were partially supported by the EU Horizon Europe MSCA grant No 101183077.

**Appendix A**

We collect basic properties of graphs and discrete dynamical systems defined on graphs. The details and proofs can be found in (Sternberg, 2010).

The matrix M is positive (M>0) if $M_{ij} > 0$ for all indices i, j. The non-negative (M≥0) is defined in the similar way. W can extend these definitions to the vectors. We say that M is primitive if there is a k such that all entries of $M^k$ are positive. M is irreducible if for any i, j there is an power k depending on these indices such that $(M^k)_{ij}>0$. Note that k can be different for different indices. We have the important theorem

**Theorem A.1 [Perron-Frobenius]**

*Let M be an irreducible matrix. Then:*

1. *M has positive eigenvalue $\lambda_{max}$ such that all other eigenvalues $\lambda$ of M satisfy $|\lambda|\leq\lambda_{max}$.*
2. *$\lambda_{max}$ has algebraic and geometric multiplicity one, and has an eigenvector v with all positive entries.*
3. *Any non-negative eigenvector is a multiple of v.*
4. *If a vector $u\geq 0$, $u\neq 0$, and $\mu$ is a number such that $Mu\leq \mu u$, then $u>0$ and $\mu\geq \lambda_{max}$.*
5. *If $0\leq S\leq M$, $S\neq M$ then every eigenvalue $\sigma$ of S satisfies $|\sigma|< \lambda_{max}$.*
6. *In particular, all diagonal minors of M have eigenvalues with absolute value less than $\lambda_{max}$.*
7. *If M is primitive, then all other eigenvalues $\lambda$ of T satisfy $|\lambda|<\lambda_{max}$.*

The theorem is important since if the TKTG is dense, then its adjacency matrix is irreducible or primitive and we can get information about its spectral properties.

The adjacency matrix M, when multiplied by itself describes a a path in the corresponding graph, e.g., $M^2p_1$ represents all vertices obtained from $p_1$ by making two jumps along the arrows.

Moreover we have that, if adjacency matrix M is primitive then, by definition, there is such k that $M^k>0$. This means that every vertex in the graph can be joined by the other one using path of length k, so the graph is strongly connected. In particular, there are cycles of length k.

For primitive matrix M, according to the Perron-Frobenius theorem there exits a maximal eigenvalue $\lambda_{max}$. and associated right x, and left y eigenvectors of M, i.e.,

$$Mx=\lambda_{max}x$$
$$yM=\lambda_{max}y’$$

and normalize as xy=1. We can construct the projector $H=x\otimes y^{+¿¿}$, that fulfills $H^2=H$. Then we have

**Theorem A.2**

$$\lim_{k\to\infty}\left(\frac{1}{\lambda_{max}}M\right)^k=H.$$

Theorem shows that the eigenvector associated with the maximal eigenvalue is a limiting point of the iteration of the matrix $\left(\frac{1}{\lambda_{max}}M\right)$.

Assume that M has only entries 0 and 1. Then we can define eigenvector centrality of the graph associated to primitive M. Consider the eigenvector x associated with maximal eigenvalue. If we normalize it $x^2=1$, then the centrality score of a vertex I is given by $x_i$. The higher score of the vertex means the higher inward edges to the node according to

$$x_i=\frac{1}{\lambda_{max}}\sum M_{ki},$$

where I(v) are indices of neighbour nodes pointing into v.

**Appendix B**

In this appendix we discuss the basics of Page Rank algorithm that is used by Google to select the best webpage resulting from search. Details can be found in (Sternberg, 2010) and (Langville, Meyer 2011). The original algorithm operates on adjacency matrix that counts the (integer) number of links between nodes representing webpages. There are extensions to arbitrary weighted graphs, see the Weighted PageRank algorithm (see Xing and Ghorbani, (2004)).

The algorithm scores the vertices of a directed graph in the situation when the adjacency matrix is not necessary a primitive one. The score of a vertex is positively influenced by the number of neighbor vertices pointing to this vertex and decreased by the outgoing vertices from the vertex, namely,

$$s(v_i) = \sum \frac{s(v_k)}{|v_i|},$$

where s(v) is a a score of the vertex v, I(v) are the indices of the neighbor vertices pointing into v, and |v| denotes the number of edges pointing out of v.

The general idea consists of amending the adjacency matrix M, that does not have to be primitive, to the primitive Google matrix G, and then using Perron-Frobenius theorem to say that there exists the left eigenvector s to the unit eigenvalue s = sG, that has components being the scores.

The eigenvectors can be obtained in iteration procedure (Sternberg, 2010) and (Langville, Meyer 2011) that reduces the complexity of computations.

**Appendix C**

In the Fourier law of thermal conduction, material thermal conductivity is defined as a proportionality constant that relates the temperature gradient to the heat flux. In our model, the analogy is not fully used, as we only need a gradient to define TKTG. We do not wish to pursue this analogy in complete detail; however, at this point, we provide an outline of the theoretical development in this direction. This Appendix can be treated as a theoretical result that expands the analogy between knowledge transfer and heat diffusion. It can serve as a foundation for future, mathematically more advanced research and its empirical verification.

The derivation utilizes discrete calculus; see, e.g., (Grady and Polimeni, 2010). It aims to rewrite continuous partial differential equations in a discrete domain, such as in our case, on graphs. In the derivation below, we will utilize the abstract index notation commonly used in tensor calculus to simplify the equations. In this framework, the quantity with two indices, e.g., $a_{ij}$, represents the whole matrix.

We can define a constant $\kappa_{ij}$, call it 'knowledge conductivity' from j to i node, that connects knowledge transfer gradient with, say, 'knowledge flux' $\vec{q_{ij}}$:

$$\vec{q_{ij}} = \kappa_{ij} \nabla a_{ij}.$$

This expresses the Fourier's law (or the Fick's law in chemical diffusion). At this point, the knowledge flux was introduced only as a theoretical construct to facilitate progress, and a detailed interpretation of this latent quantity should be provided in the domain of knowledge management. We propose interpreting it as the intensity of the actual knowledge transfer process between two employees – the actual intensity of knowledge transfer.

Assuming that every employee j has some 'reservoir' of knowledge with respect to employee i, which we can describe as a function $Q_{ij}(t)$ – the knowledge which was absorbed by the employee j with respect to the employee i. If the flow of knowledge can be treated as a flow of a 'liquid'[5] we can use the equation of continuity:

$$\frac{\partial Q_{ij}}{\partial t} = \alpha_{ij} \nabla \cdot \vec{q_{ij}},$$

where $\alpha_{ij}$ is a proportionality constant between divergence of the knowledge flux and the reservoir of knowledge, and $\nabla \cdot$ represents a divergence operator[6] on TKTG acting on the matrix q. The equation express the fact that the change in time of the reservoir of knowledge is proportional to the divergence of the knowledge flux. Roughly speaking, the equation says that the speed of change of employee's competences is related to the flux of knowledge, and the employee's absorptive capacity α. As stated in the equation, the flux can increase, but also decrease the reservoir of knowledge. It would be surprising when, during the 'knowledge transfer', the reservoir of knowledge of a teacher decreased due to the transfer to the course participant. At this point, we observe a distinction between the diffusion of a physical/chemical liquid or heat and the diffusion of knowledge. We can say that in our pursuit to make a knowledge transfer similar to the physical diffusion of heat, we faced the issue that, roughly speaking, knowledge is a special liquid that 'when we share it, it multiplies'. Mathematically, to include this fact, we must impose on the above equation the condition that the time derivative of Q is non-negative. This can be simply done by replacing right hand side (RHS) $\alpha_{ij}\nabla\cdot\vec{q_{ij}} \rightarrow \left(\alpha_{ij}\nabla\cdot\vec{q_{ij}}\right)^{+max\left(\alpha_{ij}\nabla\cdot\vec{q_{ij}},0\right)}$ - the non-negative part of RHS of the continuity equation. Such replacement can also be understood as introducing an additional source term to the continuity equation, which cancels out the terms that decrease the reservoir of knowledge Q. This also indicates that knowledge is not a conservative quantity. We will not explicitly write the positive part of the RHS below, as we aim to closely follow the derivation of the diffusion equation from heat transfer. However, at the end, we must remember that the fluxes of knowledge that drain reservoirs are forbidden. Under this assumption, the later derivations in this appendix are sensible. We would also like to emphasize that in the paper, we used only the property that the gradient induces the flow (the idea from Fourier's law) to derive TKTG and derive its properties. This is a derivation at the level of transfer of knowledge between employees. In this appendix, we further develop the analogy with diffusion, and to do so, we require the concept of a knowledge reservoir. Therefore, at this point, we must include the dynamics of the reservoir, and we face the requirement that the knowledge can be seen as a non-conservative kind of liquid that is different from physical liquid. Therefore, we must make a slight correction to the continuity equation to prevent decreasing the reservoir of knowledge Q.

[5]This is an assumption that should be verified. It results from the analogy to the heat flow, and the observation that we cannot transfer more knowledge that we possess, perhaps with some additional creative modifications. However, we can modify the proposed derivation to include some not identified internal 'sources' or 'sinks' of the knowledge by adding to the equation of continuity an additional term describing net sources or sinks of knowledge. They can be attributed to the creativity of a person or their deficiency, resulting, e.g., in forgetting or misunderstanding of a given part of knowledge. However, the exact experiments should be designed to measure properties of these quantities, which is not the purpose of this paper.

[6]Here again, there are different possibilities for define divergence for knowledge flux as for the gradient of $a_{ij}$. This is an operator that takes all components of the knowledge flux and constructs from them a new quantity on a graph, analogous to the divergence of a vector in heat conduction.

Proceeding further with the analogy od the head diffusion equation derivation, it is also natural to assume that the reservoir of knowledge $Q_{ij}$ is connected with the potential[7] $V_{ij}=a_{ij}(1)b_{ij}(1)$:

[7] We use this notion to motivate the introduction of the gradient. As before, this step is also speculative and allows for arbitrary selection of the knowledge potential. As we have seen during the definition of the gradients of knowledge, different prescriptions can be given depending on the aim.

$$\frac{\partial V_{ij}}{\partial t} = \beta_{ij} \frac{\partial Q_{ij}}{\partial t},$$

where $\beta_{ij}$ is a proportionality coefficient representing the conversion between learning potential and the knowledge reservoir quantity.

In the last step, we use the definition of the knowledge flux, continuity equation, the relation between the potential and the reservoir of knowledge to get the knowledge analogy to the heat transfer equation (assuming that β is invertible):

$$\frac{\partial a_{ij} b_i}{\partial t} = \frac{\alpha_{ij} \kappa_{ij}}{\beta_{ij}} \nabla^2 a_{ij}.$$

To get even closer resemblance to the heat/diffusion equation, we can redefine the time derivative on a graph as $\partial_t a_{ij} b_{ij} \rightarrow \partial_t a_{ij}$, to get

$$\partial_t a_{ij} = \tau_{ij} \nabla^2 a_{ij},$$

where $\tau_{ij} = \frac{\alpha_{ij} \kappa_{ij}}{\beta_{ij}}$. The Laplacian $\nabla^2$ is a linear operator resulting from the choice of the definition of a gradient and a divergence on TKTG. We can call this equation the Knowledge Diffusion Equation. It is a straightforward result when one assumes that the model of tacit knowledge is similar to the heat transfer or diffusion, and the knowledge flow resembles fluid flow.

We want to emphasize that this derivation is almost identical to the derivation of the heat equation (on a graph) using the Fourier law of heat conduction, e.g., (Grady Polimeni, 2010). To make this derivation sensible, we provided the meaning of the 'latent' variables $\vec{q}, Q$. We must also propose the interpretation for the coefficients $\alpha, \beta, \kappa, \tau$ for each pair of nodes of TKTG:

- α – *knowledge absorptive capacity* – defines how fast an employee absorbs the knowledge flux during the learning session, e.g., from words/data/learning materials; It can be used to define the learning abilities of an employee.

- β – *quality/density of knowledge* – describe how difficult or worthwhile the knowledge is. The knowledge potential V represents the will to transfer/intake of knowledge, and the coefficient β recompute it to the real competence developed by absorbing the knowledge.

- τ – *diffusion coefficients* - describe how fast the knowledge is transferred in the organization. The high value of these coefficients can indicate 'learning organization', and low value an organization based on a silos structure.

- κ – *knowledge conductivity* – can be interpreted as a throughput of the knowledge transfer channel between two employees. Its high value can indicate employees who understand each other implicitly. Low values can indicate a mismatch, conflicts, or bureaucratic barriers.

One can note that the matrices $\alpha, \beta, \kappa, \tau$ for the heat equation are connected, as for the diffusion/heat equation, with the fundamental properties of the 'knowledge conducting material' in the company. Therefore, these coefficients can be used to characterize the 'material', attributed to the ease of knowledge transfer, speed, and resistance of this transfer in the company between pairs of employees (and domains represented by them). We refer to the detailed characterization of them for future research in this direction. The main emphasis will be to measure the latent variables Q and q effectively, from which we can determine 'material coefficients'.

We also want to remark that the diffusion equation, as an example of parabolic/diffusion Partial Differential Equations, has a property of 'smoothing' of initial data (Robinson, 2010). Roughly speaking, if there is an excess or a lack of fluid/heat/substance in some regions at the beginning, during the diffusion, the fluid/heat/substance will be spread evenly in the whole region. Similarly, with knowledge, if sufficient conditions (set up by $a_{ij}$, $b_{ij}$ at the initial time) for transfer are met in the company, then the excess or lack of knowledge[8], via learning sessions, should be 'equalized' among employees in a the long run. We therefore sketched the proof of the following:

**Theorem C.1**

*The approach of organizing knowledge exchange sessions based on TKTG leads to equalizing/diffusion of knowledge among the learning groups members.*

The above theorem expresses what we intuitively expect from the process of knowledge transfer and confirms that the proposed model of TKTG is a correct model for knowledge transfer.

**References**

1. Ahmad, R. , Azman Ali, N. (2003). *The use of cognitive mapping technique in management research: theory and practice,* Management Research News, Vol. 26 No. 7
2. Augier, M., Vendelo, M.T. (1999). *Networks, cognition and management of tacit knowledge,* Journal of Knowledge Management, Vol. 3 No. 4

[8]We want to emphasize that only transfers that do not decrease the reservoir are allowed in the end. Therefore, the employees with a lack of knowledge will increase their reservoir of knowledge. On the other hand, employees with an excess of knowledge will still possess it after the correction. Therefore, the overall knowledge in the company will grow.

3. Baumard, P. (2002). *Tacit knowledge in professional firms: the teachings of firms in very puzzling situations,* Journal of Knowledge Management, Vol. 6 No. 2
4. Bergman, J.-P., Knutas, A., Luukka, P., Jantunen, A., Tarkiainen, A., Karlik, A. and Platonov, V. (2016), *Strategic interpretation on sustainability issues – eliciting cognitive maps of boards of directors,* Corporate Governance, Vol. 16 No. 1
5. Binney, J. J., Dowrick, N. J., Fisher, A. J., Newman, M. E. J. (1992). *The Theory of Critical Phenomena: An Introduction to the Renormalization Group*, Clarendon Press
6. Boamah, F.A., Zhang, J., Miah, M.H. (2023). *The impact of tacit knowledge sharing on the success of construction companies operations,* Journal of Engineering, Design and Technology, Vol. 21 No. 6
7. Brockmann, E.N., Anthony, W.P. (1998). *The influence of tacit knowledge and collective mind on strategic planning,* Journal of Managerial Issues, Vol. X No. 2
8. Cen, D., Teichert, E., Hodgetts, C.J. et al. (2024). *Curiosity shapes spatial exploration and cognitive map formation in humans,* Communications Psychology Vol 2, No 129. https://doi.org/10.1038/s44271-024-00174-6
9. Cengel Y.A.,, Ghajar, A.J. (2019). *Heat and Mass Transfer: Fundamentals and Applications*, McGraw Hill; 6th edition
10. Chen, L., Mohamed, S. (2010). *The strategic importance of tacit knowledge management activities in construction,* Construction Innovation, Vol. 10 No. 2
11. Daga, H., Srivastava, Ch., Choudhary, L. (2025). *The economics of tacit knowledge in patent based technology transfer agreements,* The Journal of World Intellectual Property. 26 September 2025 https://doi.org/10.1111/jwip.70007
12. Evariste, M. (2019). *A review on rockburst as a serious safety problem in deep underground mines and other excavation projects,* International Journal of Engineering Research and Technology Volume 8 Issue 11 .
https://doi.org/10.17577/ijertv8is110194

13. Gamble, J.R. (2020). *Tacit vs explicit knowledge as antecedents for organizational change,* Journal of Organizational Change Management, Vol. 33 No. 6,
14. Gill, S.P. (2000). *The tacit dimension of dialogue for knowledge transfer,* Conference proceedings from the Conference on Knowledge and Innovation, Helsinki School of Economics and Business Administration
15. L.J. Grady, J.R. Polimeni, *Discrete Calculus. Applied Analysis on Graphs for Computational Science,* Springer-Verlag London Limited 2010
16. Haldin-Herrgard, T. (2000). *Difficulties in diffusion of tacit knowledge in organizations,* Journal of Intellectual Capital, Vol. 1 No. 4
17. Hodgkinson, G.P. (Ed.) *Methodological Challenges and Advances in Managerial and Organizational Cognition,* New Horizons in Managerial and Organizational Cognition, Vol. 2,
18. Holste, J.S., Fields, D. (2010). *Trust and tacit knowledge sharing and use*, Journal of Knowledge Management, Vol. 14 No. 1
19. Ilinski, K. (2001). *Physics of Finance: Gauge Modelling in Non-Equilibrium Pricing,* Wiley
20. Isard, W. (1954). *Location Theory and Trade Theory: Short-Run Analysis,* Quarterly Journal of Economics. 68 (2): 305–320
21. Jafari, M., Akhavan, P., Nikookar, M. (2013). *Personal knowledge management and organization's competency: a service organization case study,* Education, Business and Society: Contemporary Middle Eastern Issues, Vol. 6 No. ¾

22. Kågström, B., Ruhe, A. (1980). *An Algorithm for Numerical Computation of the Jordan Normal Form of a Complex Matrix*, ACM Transactions on Mathematical Software (TOMS), Volume 6, Issue 3, 398 – 419; DOI: https://doi.org/10.1145/355900.355912
23. Kosko, B. (1992). *Neural networks and fuzzy systems*, USA: Prentice Hall International Editions
24. Kruger, J. Dunning, D. (1999). *Unskilled and unaware of it: how difficulties in recognizing one's own incompetence lead to inflated self-assessments,* Journal of Personality and Social Psychology. 77 (6): 1121–34. doi:10.1037/0022-3514.77.6.1121
25. Krumpal, I. (2013). *Determinants of social desirability bias in sensitive surveys: a literature review,* Quality & Quantity. 47 (4): 2025–2047. doi:10.1007/s11135-011-9640-9
26. Kümmel R. (2011). *The Second Law of Economics,* Springer
27. Kwong, E., Lee, W.B. (2009), *Knowledge elicitation in reliability management in the airline industry,* Journal of Knowledge Management, Vol. 13 No. 2
28. Langville, A.N., Meyer C.D. (2011), *Google's PageRank and Beyond*, Princeton University Press
29. Lee Endres, M., Endres, S.P., Chowdhury, S.K., Alam, I. (2007). *Tacit knowledge sharing, self-efficacy theory, and application to the Open Source community,* Journal of Knowledge Management, Vol. 11 No. 3
30. Leonard, D., Sensiper, S. (1998). *The role of tacit knowledge in group innovation*. California Management Review, Vol. 40 No. 3

31. Moustaghfir, K., Schiuma, G. (2013). *Knowledge, learning, and innovation: research and perspectives*. Journal of Knowledge Management, Vol. 17 No. 4,
32. Mühlenthaler M. (2015). *Fairness in Academic Course Timetabling*, Lecture Notes in Economics and Mathematical Systems, Springer Cham
33. Nieżurawska, J., Niemczynowicz, A., Kycia, R.A. (2023). *Managing Generation Z. Motivation, Engagement and Loyalty*, Routledge New York
34. Pathirage, C.P., Amaratunga, D.G., Haigh, R.P. (2007). *Tacit knowledge and organisational performance: construction industry perspective*. Journal of Knowledge Management, Vol. 11 No. 1
35. Penrose, R., Rindler, W. (1987). S*pinors and Space-Time*, Vol. 1-2, Cambridge University Press
36. Polanyi, M. (1966). *The Tacit Dimension*. Routledge & Kegan Paul Ltd, London.
37. Powell, B.C. (2023). *Sharing intractably tacit knowledge: the case for nepotistic selection*. Development and Learning in Organizations, Vol. 37 No. 5
38. Richardson, G. (2004). *Guilds, laws, and markets for manufactured merchandise in late-medieval England*, Explorations in Economic History 41, 1-25
39. Robinson, J.C. (2010). *Infinite-Dimensional Dynamical Systems: An Introduction to Dissipative Parabolic PDEs and the Theory of Global Attractors,* Cambridge University Press
40. Rosselini, A., Hawamdeh, S. (2020). *Tacit knowledge transfer in training and the inherent limitations of using only quantitative measures.* Proceedings of the Association for Information Science and Technology: Volume 57, Issue 1
41. Rogers, E. (2003). *Diffusion of Innovations*, 5th Edition. Simon and Schuster.

42. Rosario M.N., Eugene H. (2007). *Introduction to Econophysics: Correlations and Complexity in Finance*. Cambridge University Press
43. Ruíz-Tolosa, J.R., Castillo, E. (2005). *From Vectors to Tensors*, Springer Berlin, Heidelberg
44. Shi, Q., Wang, Q., Guo, Z. (2022). *Knowledge sharing in the construction supply chain: collaborative innovation activities and BIM application on innovation performance*. Engineering, Construction and Architectural Management, Vol. 29 No. 9
45. Sternberg S. (2010). *Dynamical Systems*, Dover Publications
46. Xing, W., Ghorbani, A. (2004). *Weighted PageRank algorithm*. W: Proceedings of the Second Annual Conference on Communication Networks and Services Research (CNSR'04) pp. 305-310 IEEE.
47. Zeng, Z., Tien-Yien Li, T.-L., (2021). *A numerical method for computing the Jordan Canonical Form*, ArXiv: 2103.02086 [math.NA]